\documentclass[latin1]{aa}
\usepackage[english]{babel}
\usepackage{multirow}
\usepackage{latexsym}
\usepackage{amssymb}
\usepackage{relsize}
\usepackage{hyperref}
\usepackage{txfonts}
\usepackage{array} 
\usepackage{tablefootnote}
\bibpunct{(}{)}{,}{a}{}{,}

\newcommand{\eg}{\textit{e.g.}}
\newcommand{\drho}{\frac{\delta \rho}{\rho_0}}
\newcommand{\dP}{\frac{\delta p}{P_0}}
\newcommand{\dPi}{\delta p/P_0}
\newcommand{\vect}{\vec}
\newcommand{\vlp}{\left(\omega + m\Omega\right)}
\renewcommand{\dfrac}[2]{\frac{\mathrm{d} #1}{\mathrm{d} #2}}

\newcommand{\dpart}[2]{\frac{\partial #1}{\partial #2}}
\newcommand{\ddpart}[2]{\frac{\partial^2 #1}{\partial #2^2}}
\newcommand{\grad}{\vect{\nabla}}

\newcommand{\lapl}{\Delta}
\newcommand{\Req}{R_{\mathrm{eq}}}
\renewcommand{\div}{\vect{\nabla} \cdot}
\renewcommand{\l}{\ell}

\newcommand{\rz}{r_{\zeta}}
\newcommand{\rt}{r_{\theta}}
\newcommand{\rzz}{r_{\zeta\zeta}}
\newcommand{\rzt}{r_{\zeta\theta}}
\newcommand{\rtt}{r_{\theta\theta}}
\newcommand{\Gammam}{{\Gamma\!\!\scriptscriptstyle -}}
\newcommand{\Gammap}{{\Gamma\!\!\scriptscriptstyle +}}

\renewcommand{\d}{\partial}

\newcommand{\dz}{\partial_{\zeta}}
\newcommand{\dt}{\partial_{\theta}}
\newcommand{\dphi}{\partial_{\phi}}
\newcommand{\dzz}{\partial_{\zeta\zeta}^2}
\newcommand{\dzt}{\partial_{\zeta\theta}^2}
\newcommand{\dtt}{\partial_{\theta\theta}^2}

\newcommand{\gzz}{g^{\zeta\zeta}}

\newcommand{\xiz}{\xi^{\zeta}}
\newcommand{\xit}{\xi^{\theta}}
\newcommand{\xip}{\xi^{\phi}}

\newcommand{\sint}{\sin\theta}
\newcommand{\cost}{\cos\theta}
\newcommand{\cott}{\cot\theta}
\newcommand{\xT}{x_{\mathrm{T}}}
\newcommand{\tauT}{\tau_{\mathrm{T}}}

\newcommand{\kt}{k_{\theta}}
\newcommand{\vkpar}{\vect{k}_{\parallel}}
\newcommand{\kpar}{k_{\parallel}}
\newcommand{\kperp}{k_{\perp}}
\newcommand{\ez}{\vect{e}_z}
\newcommand{\OmegaK}{\Omega_{\mathrm{K}}}

\newcommand{\Nr}{N_{\mathrm{r}}}
\newcommand{\Nrmod}{N_{\mathrm{r,\,mod}}}
\newcommand{\Nt}{N_{\theta}}

\begin{document}
   \title{Oscillations of 2D ESTER models}
   \subtitle{I. The adiabatic case}

   \author{D. R. Reese\inst{1} \and G. M. Mirouh\inst{2} \and F. Espinosa Lara\inst{3} \and
           M. Rieutord\inst{4,5} \and B. Putigny\inst{4,5}}

   \institute{
           LESIA, Observatoire de Paris, Université PSL, CNRS,
           Sorbonne Université, Univ. Paris Diderot, Sorbonne Paris Cité,
           5 place Jules Janssen, 92195 Meudon, France \\
           \email{daniel.reese@obspm.fr}
           \and
           Astrophysics Research Group, Faculty of Engineering 
and Physical Sciences, University of Surrey, Guildford GU2 7XH, UK
           \and
           Universidad de Alcalá, Space Research Group, 28805 Alcalá de Henares, Spain
           \and
           Universt{\'e} de Toulouse, UPS-OMP, IRAP, Toulouse, France
           \and
           CNRS, IRAP, 14 avenue Edouard Belin, 31400 Toulouse, France
   }

   \date{}

  \abstract
   {Recent numerical and theoretical considerations have shown that low-degree
   acoustic modes in rapidly rotating stars follow an asymptotic formula. In
   parallel, recent studies have revealed the presence of regular pulsation
   frequency patterns in rapidly rotating $\delta$ Scuti stars that seem to
   match theoretical expectations.}
   {In this context, a key question is whether strong gradients or
   discontinuities can adversely affect the asymptotic frequency pattern to the
   point of hindering its identification.  Other important questions are how
   rotational splittings are affected by the 2D rotation profiles expected from
   baroclinic effects and whether it is possible to probe the rotation profile
   using these splittings.}
   {In order to address these questions, we numerically calculate stellar
   pulsation modes in continuous and discontinuous rapidly rotating models
   produced by the 2D ESTER (Evolution STEllaire en Rotation) code.  This code
   self-consistently calculates the rotation profile based on baroclinic effects
   and uses a spectral multi-domain approach, thus making it possible to
   introduce discontinuities at the domain interfaces without loss of numerical
   accuracy.  The pulsation calculations are carried out using an adiabatic
   version of the Two-dimensional Oscillation Program (TOP) code.  The
   variational principle is then used to confirm the high numerical accuracy of
   the pulsation frequencies and to derive an integral formula for the
   generalised rotational splittings.  Acoustic glitch theory, combined with ray
   dynamics, is applied to the discontinuous models in order to interpret their
   pulsation spectra.}
   {Our results show that the generalised rotational splittings are very well
   approximated by the integral formula, except for modes involved in avoided
   crossings.  This potentially allows the application of inverse theory for
   probing the rotation profile.  We also show that glitch theory
   applied along the island mode orbit can correctly predict the periodicity of
   the glitch frequency pattern produced by the discontinuity or $\Gamma_1$ dip
   related to the He II ionisation zone in some of the models.  Furthermore,
   the asymptotic frequency pattern remains sufficiently well preserved to
   potentially allow its detection in observed stars.}
   {}

   \keywords{stars: oscillations (including pulsations) -- stars: rotation -- stars: interiors}

   \maketitle
%

\section{Introduction}

Much effort has gone into producing realistic models of rapidly rotating stars.
This includes the pioneering works by \citet{Roxburgh1965, Ostriker1968}, and
\citet{Jackson1970} and continues on in the present with various 1D codes
\citep[\eg][]{Palacios2003,Eggenberger2008,Marques2013} as well as 2D codes such
as the one from the Evolution STEllaire en Rotation (ESTER) project
\citep{Rieutord2009, EspinosaLara2013, Rieutord2016}.  An extensive monograph on
the effects of rotation on stellar structure and evolution has also recently
been published \citep{Maeder2009}. In parallel, much work has gone into
calculating pulsation spectra in such models in order to interpret observations
from recent space missions such as CoRoT \citep{Baglin2009, Auvergne2009},
Kepler \citep{Borucki2009}, and TESS \citep{Ricker2015}.  Some of the most
recent works include \citet{Lovekin2008, Lovekin2009, Lignieres2008,
Lignieres2009, Ballot2010, Reese2009a, Reese2013, Ouazzani2015}, and
\citet{Ouazzani2017}.  Of these works, only \citet{Ouazzani2015} addresses
pulsations in baroclinic stellar models, that is, models in which surfaces of
constant pressure, temperature, or density do not coincide. This is a major
ingredient of realistic models, as rotating stars are expected to be baroclinic
\citep[\eg][]{Zahn1992}.  The work by \citet{Ouazzani2015} used stellar models
from \citet{Roxburgh2006} in which the rotation profile is imposed beforehand
rather than being calculated in a self-consistent way using energy conservation.
In contrast, the ESTER code deduces the rotation profile in a self-consistent
way when constructing stellar models.  Hence, it is important to study pulsation
modes in such models.

One of the first signatures of rotation on stellar pulsations is rotational
splittings, the frequency differences between consecutive modes with the same
radial order and harmonic degree but different azimuthal orders.  At slow
rotation rates, rotational splittings can be used to invert 1D or 2D rotation
profiles using a first-order perturbative approach \citep[\eg][]{Deheuvels2014,
Schou1998, Thompson2003}.  At high rotation rates, higher-order effects come
into play and must be addressed before meaningful information on the rotation
profile can be deduced \citep[\eg][]{Soufi1998, Suarez2009}.  In this context, a
particularly interesting quantity to investigate is the generalised rotational
splitting, namely the frequency difference between prograde modes and their
retrograde counterparts.  In particular, \citet{Ouazzani2012b} showed that it is
possible to distinguish between third-order effects of rotation and latitudinal
differential rotation in such splittings.  At higher rotation rates,
\citet{Reese2009a} showed that such splittings are a weighted integral of the
rotation profile, provided the degree of differential rotation is not too
large.  This would potentially provide the basis for carrying out rotation
inversions in such stars.  This work, however, was restricted to cylindrical
rotation profiles and furthermore neglected the influence of the Coriolis force
in the integrals.  This raises the open questions of whether such weighted
integrals can be generalised to general 2D rotation profiles, and if so, how
accurate they are.

Another important consideration concerns frequency separations.  Indeed, a
number of recent studies have shown that the pulsation frequencies of low-degree
acoustic modes of rapidly rotating stars follow an asymptotic formula.  Such a
formula was first explored on an empirical basis \citep{Lignieres2006,
Reese2008a, Reese2009a} before being justified using ray dynamics
\citep{Lignieres2008, Lignieres2009, Pasek2011, Pasek2012}.  \citet{Reese2017}
studied theoretical pulsation spectra with realistic mode visibilities in
rapidly rotating $1.8$ and $2$ M$_{\odot}$ stellar models based on the
self-consistent field (SCF) method \citep{Jackson2005, MacGregor2007}. They
showed that it may be possible, depending on the configuration, to detect the
rotating counterpart to the large frequency separation, or half its value, as
well as frequency spacings corresponding to multiples of the rotation rate. 
More recently, \citet{Mirouh2019} set up a machine learning algorithm to
automatically identify to which class a given mode belongs.  They went
on to characterise the large frequency separation in a large set of models at
different rotation rates and with different core compositions (thus mimicking
the effects of stellar evolution), and showed a tight scaling relation between
it and the stellar mean-density. From an observational point of view, recurrent
frequency spacings have been detected in a number of $\delta$ Scuti stars
\citep{Mantegazza2012, Suarez2014, GarciaHernandez2009, GarciaHernandez2013,
Paparo2016}, including a very recent study involving interferometry,
spectroscopy, and space photometry \citep{Bouchaud2020}, and interpreted as the
large frequency separation or half its value.  \citet{GarciaHernandez2015}
studied a number of $\delta$ Scuti pulsators in binary systems, for which
independent estimates of the mass and radius are available, and have shown that
this separation scales with the mean density, as expected based on the
calculations in \citet{Reese2008a}. Ensemble asteroseismology has
recently been applied to CoRoT $\delta$ Scuti stars by \citet{Michel2017}
who also found regular patterns related to the large separation, although we
note that \citet{Bowman2018} applied a similar strategy to Kepler $\delta$ Scuti
stars without the same degree of success.  Finally, in the very recent work by
\citet{Bedding2020}, the pulsation spectra of 57 $\delta$ Scuti stars observed
by TESS and three by Kepler were matched to axisymmetric $\l=0$ and $\l=1$ modes
from non-rotating models via echelle diagrams.  Such modes were shown
to be relatively invariant as a function of rotation rate up to $\sim
0.5\,\OmegaK$ using pulsation calculations in SCF models, apart from a scale
factor related to the mean density, thus justifying the use of non-rotating
models.

However, it is unclear to what extent the asymptotic formula would hold in the
presence of discontinuities within the stellar model.  Based on results
previously obtained in non-rotating models with sharp gradients
\citep[\eg][]{Monteiro1994}, one can expect the asymptotic formula to still
apply albeit with a supplementary oscillatory component.  However, it is not
clear how strong this component is, how it behaves in the presence of rapid
rotation, and whether it can hinder the interpretation of observed oscillation
spectra in rapidly rotating stars, as discussed in \citet{Breger2012}.  The
recent works by \citet{Bouabid2013} and \citet{Ouazzani2017} have shown, using
the traditional approximation and full 2D pulsation calculations, respectively,
how a sharp gradient around the core of rotating $\gamma$ Dor stars affects
g-modes.  In particular, they show the presence of a periodic component in the
period spacings of such modes, analogous to what was found in the non-rotating
case \citep{Miglio2008}, in agreement with observations from the Kepler mission
\citep[\eg][]{VanReeth2015}.  Likewise, similar observations in SPB stars have
also revealed an oscillatory behaviour in the period spacing of their g-modes
\citep[\eg][]{Papics2017}.  A similar study is needed for acoustic modes in
rapidly rotating stars.

In order to address the above questions, we investigate low-degree acoustic
modes in rapidly rotating stellar models from the ESTER code.  One of the
advantages of the ESTER code is its multi-domain spectral approach, ideal for
introducing discontinuities while retaining a high numerical accuracy.  The
pulsation modes are calculated using a multi-domain spectral version of the
Two-dimensional Oscillation Program \citep[TOP,][]{Reese2006, Reese2009a}.  The
article is organised as follows: the following section describes stellar models
based on the ESTER code.  This is then followed by a description of the
pulsation calculations as well as the variational principle, with a particular
emphasis on the effects of discontinuities.  Section~\ref{sect:rotation} deals
with generalised rotational splittings.  Section~\ref{sect:glitches} then goes on
to describe the effects of discontinuities, both on the pulsation frequencies
and on the eigenmodes. This is then followed by the conclusion.

\section{Stellar models based on the ESTER code}

The aim of the ESTER project is to produce and evolve self-consistent stellar
models of rapidly rotating stars.  Consequently, a fully 2D approach is used in
order to solve the relevant fluid equations while taking into account energy
conservation when modelling the stationary structure of the star.  This leads to
centrifugal deformation of the stellar structure, as well as more subtle
effects, namely differential rotation and meridional circulation, resulting from
baroclinicity.  Consequently, the rotation profile depends on both the radial
coordinate and colatitude, and the isobars, isochores, and isotherms are
distinct.

In terms of microphysics, it is possible to apply various equations of state
(EOS) in ESTER.  These include: the ideal gas law with or without radiation
pressure, the OPAL EOS \citep{Rogers2002}, and
FreeEOS\footnote{\url{http://freeeos.sourceforge.net/}} \citep{Irwin2012}.  In
what follows, we applied the ideal gas law (without radiation pressure) in the
discontinuous models (see below) and one of the continuous models, in order to
avoid introducing numerical errors coming from a tabulated EOS, and the OPAL EOS
in the other continuous model for the sake of realism.  In terms of opacities,
there are two options currently implemented: Kramer's opacities and OPAL
opacities \citep{Iglesias1996}.  We used Kramer's opacities in conjunction with
the ideal gas law to rely entirely on analytical expressions thus reducing
numerical errors, and OPAL opacities with the OPAL EOS for the sake of realism
and consistency.  Models with Kramer's opacity have significantly larger radii
and hence lower mean densities.

Currently, the ESTER code has some limitations.  Firstly, it is unable to
simulate convective envelopes.  Indeed, applying a strong entropy diffusion as
is done in the convective core is too approximate for the envelope.  Various
numerical difficulties have so far prevented the code from converging to a
convective solution in such regions.  Accordingly, ESTER is currently not
suitable for stars with masses below $\sim 1.6$ M$_{\odot}$.  Secondly, the
ESTER code is unable to simulate time evolution using a full chain of nuclear
reactions.  However, it is possible to alter the core composition in order to
mimic the effects of stellar evolution or to include a rudimentary
implementation of hydrogen combustion.

From a numerical point of view, the star is divided into multiple domains in the
radial direction.  There are two main reasons for doing this.  First, this
allows us to overcome the limitations inherent to using a spectral approach with
its imposed collocation grid.  In particular, it enables us to have a high
resolution near the surface where it is needed.  The second reason is that one
can place a discontinuity between two domains without losing spectral accuracy.
This goes hand in hand with the use of a dedicated coordinate system,
$(\zeta,\theta,\phi)$, where $\zeta$ is a surface-fitting radial coordinate that
is constant across the stellar surface and across the surfaces which delimit the
boundary between consecutive domains \citep[see][for more
details]{Rieutord2016}.

\begin{figure}
\includegraphics[width=\columnwidth]{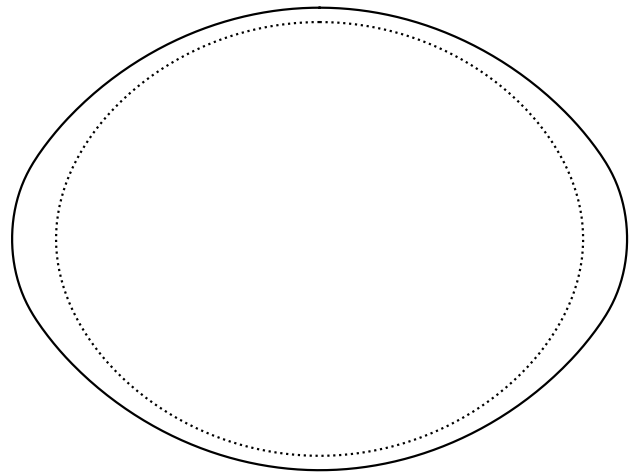}
\caption{Meridional cross-section of model \texttt{M6} showing where the
discontinuity is located.  The other models have a discontinuity closer to
the surface. \label{fig:discontinuity_2D}}
\end{figure}

\begin{figure}
\includegraphics[width=\columnwidth]{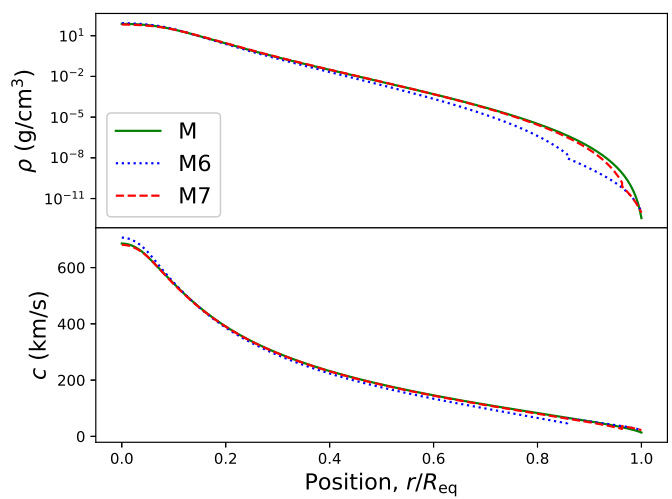}
\caption{Density (upper panel) and sound velocity (lower panel) profiles 
along the equator in models \texttt{M}, \texttt{M6}, and \texttt{M7}.
\label{fig:discontinuity_profiles}}
\end{figure}

In this study, we use 2 $M_{\odot}$ stellar models at $70 \%$ of the Keplerian
break-up rotation rate.  We note that this value is not too far from the
rotation rates of Rasalhague ($\alpha$ Oph) for which $\Omega \sim 0.64 \OmegaK$
\citep[see \eg][and references therein]{Deupree2011, Mirouh2017} and Altair for
which $\Omega = 0.74 \OmegaK$ \citep{Bouchaud2020}, two well-studied $\delta$
Scuti stars with photometric observations from the space missions MOST and WIRE
respectively. These models use a spectral approach based on Chebyshev
polynomials in the radial direction, and spherical harmonics in the horizontal
directions.  The radial direction is subdivided into eight domains, the
resolution in each domain being 30, 55, 45, 40, 40, 50, 70, and 70, that is, a
total of 400 radial points.  In the horizontal directions, 22 or 32 points on
half of a Gauss-Legendre collocation grid are used depending on the model, thus
corresponding to 22 or 32 spherical harmonics with even $\l$ values.  Density
discontinuities are achieved by modifying the hydrogen content abruptly. 
Table~\ref{tab:characteristics} gives the characteristics of the five models
(\texttt{Mreal}, \texttt{M}, \texttt{M6}, \texttt{M7}, and \texttt{M7b})
involved in this study. In all of the models, $Z = 0.02$ everywhere.
Figure~\ref{fig:discontinuity_2D} shows where the discontinuity is located in
model \texttt{M6}, and Fig.~\ref{fig:discontinuity_profiles} gives the density
and sound velocity profiles in the \texttt{M}, \texttt{M6}, and \texttt{M7}
models.

\begin{table*}[h!]
\begin{center}
\caption{Characteristics of the models used in this study.  $X_{\mathrm{int}}$
and $X_{\mathrm{ext}}$ are the hydrogen contents below and above the
discontinuity, respectively. $R_{\mathrm{disc}}/R_{\mathrm{eq}}$ gives the
equatorial radius at the discontinuity, normalised by the star's equatorial
radius.  We note that models \texttt{Mreal} and \texttt{M} are smooth.
\label{tab:characteristics}}
\begin{tabular}{cccccccc}
\hline
\textbf{Model name} &
$X_{\mathrm{int}}$ &
$X_{\mathrm{ext}}$ &
$R_{\mathrm{disc}}/R_{\mathrm{eq}}$ &
$\rho_0^+/\rho_0^-$ &
$c_0^+/c_0^-$ &
\textbf{EOS} &
\textbf{Opacities} \\
\hline
\texttt{Mreal}  & 0.70 & 0.70 & n.a. & 1 & 1 & OPAL & OPAL \\ 
\texttt{M}      & 0.70 & 0.70 & n.a. & 1 & 1 & Ideal gas & Kramer \\
\texttt{M6}     & 0.07 & 0.70 & 0.857074 & 0.513889 & 1.394972 & Ideal gas & Kramer \\ 
\texttt{M7}     & 0.07 & 0.70 & 0.962678 & 0.513889 & 1.394972 & Ideal gas & Kramer \\ 
\texttt{M7b}    & 0.70 & 0.07 & 0.991861 & 1.945946 & 0.716860 & Ideal gas & Kramer \\ 
\hline
\end{tabular}
\end{center}
\end{table*}

Models \texttt{Mreal} and \texttt{M} are our most realistic models, and serve as
a reference, since they do not feature ad hoc discontinuities.  Models
\texttt{M6} and  \texttt{M7} include a drop in density near the surface for two
different radii, while \texttt{M7b} includes an increase in density near the
surface.  In realistic models, such as \texttt{Mreal}, such discontinuities are
not expected.  Instead, more subtle phenomena, such as dips in the $\Gamma_1$
profile due to the hydrogen and helium ionisation zones, occur near the stellar
surface and can lead to a glitch pattern in the frequencies.  However, it is
still useful to test models with discontinuities as they exaggerate the
phenomena we wish to study, namely acoustic glitches, and should thus make it
easier to detect its signature in the pulsation spectrum.  Furthermore, one can
easily modify the different parameters related to the discontinuity such as
depth and intensity in order to study its impact on the frequencies. Finally,
the lack of radiation pressure in these models leads to flat $\Gamma_1$ profiles
meaning that the only glitch signatures expected are those arising from the
discontinuities, thus simplifying the subsequent analysis. Nonetheless, the
realistic model also allows us to test acoustic glitches related to the
$\Gamma_1$ profile.

We do note that stars can have a discontinuity around the core due to
the depletion of hydrogen by nuclear reactions. However, acoustic modes
are sensitive to the near-surface layers of stars and are thus not the most
suitable for studying such discontinuities.  This is particularly true of
island modes as the ray trajectory orbits around which they are concentrated
remain far away from the convective core for $\Omega \gtrsim 0.2\,\OmegaK$ (at
least for the models in this study).  In order to probe such discontinuities, it
is more useful to look at gravity-mode glitches \citep[\eg][]{Miglio2008,
Ouazzani2017}, which is beyond the scope of the present article.

\begin{figure}[htbp]
\includegraphics[width=\columnwidth]{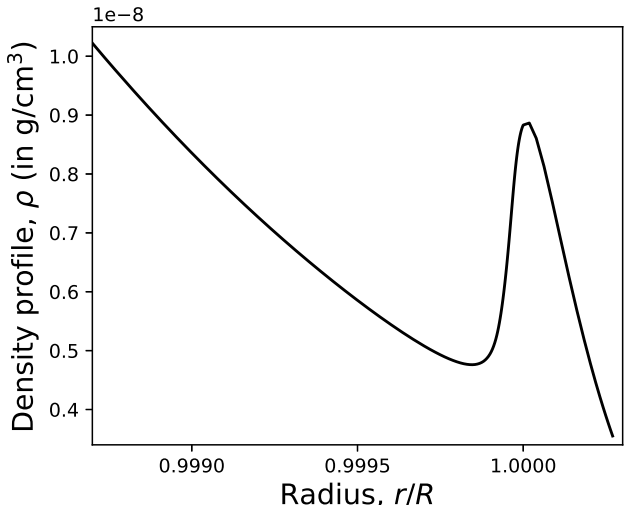}
\caption{Near-surface density profile in a $2$ M$_{\odot}$ non-rotating main
sequence model from grid B of \citet{Marques2008}. \label{fig:density_inversion}}
\end{figure}

In model \texttt{M7b}, the denser layer is on top.  At first sight, this
may seem unrealistic, but density inversions can occur in the near-surface
layers of stars such as the one shown in Fig.~\ref{fig:density_inversion} for a
$2$ M$_{\odot}$ non-rotating main sequence model from grid B of
\citet{Marques2008}.  Such density inversions typically occur as a result of a
sharp temperature drop in low density regions of the star (Marques, private
communication).  Furthermore, including a model with a denser layer on top
allows us to test the Snell-Descartes law in different situations.

\section{Pulsation calculations}

The pulsation modes are calculated using the Two-dimensional Oscillation
Program \citep[TOP,][]{Reese2006, Reese2009a}.  This program fully takes into
account the centrifugal deformation and has been set up to apply a multi-domain
spectral approach, in accordance with the models from ESTER. The next
subsections describe the set of pulsation equations, the interface conditions
that apply between different domains, the boundary conditions, and the
numerical approach.

\subsection{Pulsation equations}
The following set of equations are used to calculate pulsation modes.  They are,
respectively, the continuity equation, Euler's equation, the adiabatic relation,
and Poisson's equation:
\begin{eqnarray}
\label{eq:continuity}
0 &=& \drho + \div \vect{\xi}, \\
\label{eq:Euler}
0 &=& \vlp^2 \vect{\xi} - 2i \vlp \vect{\Omega} \times \vect{\xi}
      - \vect{\Omega} \times \left( \vect{\Omega} \times \vect{\xi} \right) \nonumber \\
  & & - \vect{\xi} \cdot \grad \left( s \Omega^2 \vect{e}_s\right)
      - \frac{P_0}{\rho_0} \grad \left(\dP\right)
      + \frac{\grad P_0}{\rho_0} \left(\drho - \dP\right) - \grad \Psi \nonumber \\
  & & + \grad \left(\frac{\vect{\xi}\cdot\grad P_0}{\rho_0}\right)
      + \left[\frac{\left(\vect{\xi}\cdot\grad P_0\right)\grad\rho_0 -
            \left(\vect{\xi}\cdot\grad\rho_0\right)\grad P_0}{\rho_0^2}\right], \\
\label{eq:energy}
0 &=& \dP - \Gamma_1 \drho, \\
\label{eq:Poisson}
0 &=& \lapl \Psi - \Lambda \left(\rho_0\drho - \vect{\xi}\cdot\grad\rho_0\right),
\end{eqnarray}
where quantities with the subscript `0' are equilibrium quantities, those with
`$\delta$' in front Lagrangian perturbations, $\rho$ the density, $P$ the
pressure, $\Psi$ the Eulerian gravitational potential perturbation, $\vect{\xi}$
the Lagrangian displacement, $\Omega$ the rotation profile (which depends on
$\zeta$, the surface-fitting radial coordinate, and $\theta$, the
co-latitude), $\Lambda = 4\pi G$, $G$ the gravitational constant, $s$ the
distance from the rotation axis, and $\vect{e}_s$ the associated unit vector.
The term in square brackets (last line of Eq.~(\ref{eq:Euler})) does not cancel,
since the stellar model is not barotropic.  The Lagrangian density perturbation
is eliminated in favour of the Lagrangian pressure perturbation using
Eq.~(\ref{eq:energy}).  In the above set of equations, we have neglected the
meridional circulation, given that it is expected to have a negligible effect on
the pulsation modes.

\subsection{Non-dimensionalisation}

The following reference length, pressure, and density scales are used:
\begin{equation}
R_{\mathrm{ref}} = \Req, \qquad P_{\mathrm{ref}} = \frac{GM^2}{\Req^4},\qquad
\rho_{\mathrm{ref}} = \frac{M}{\Req^3},
\end{equation}
where $\Req$ is the equatorial radius and $M$ the mass.  As a result
of this choice of reference scales, the frequencies are non-dimensionalised
by the inverse of the dynamic time scale:
\begin{equation}
\omega_{\mathrm{ref}} = \OmegaK = \sqrt{\frac{GM}{\Req^3}}.
\end{equation}
Using this non-dimensionalisation leads to the same set of pulsation equations
as previously (Eqs.~(\ref{eq:continuity})-(\ref{eq:Poisson})) except that
$\Lambda$ is now equal to $4\pi$.

\subsection{Interface conditions}

Given that ESTER models are calculated over multiple domains, interface
conditions are needed to describe the relation between various quantities on
either side of the different boundaries.  Furthermore, care is needed when
expressing these conditions given that some of the models contain
discontinuities. The first condition is simply that the fluid domain is
continuous. In other words, the deformation of the boundary must be the same on
either side. This yields the following first order expression:
\begin{equation}
\label{eq:interface_displacement}
\vect{\xi}_- \cdot \vect{n} = \vect{\xi}_+ \cdot \vect{n},
\end{equation}
where $\vect{n}$ is the normal to the unperturbed surface, and the subscripts
`-' and `+' denote quantities below and above the boundary.  This condition does
allow the fluid to `slip' along the boundary.  A more detailed derivation is
given in App.~\ref{app:interface_displacement}.

A second condition is that the pressure remains continuous across the perturbed
boundary.  This condition is simply expressed as follows (see
App.~\ref{app:interface_pressure}):
\begin{equation}
\label{eq:interface_pressure}
\delta p_- = \delta p_+.
\end{equation}

The third condition is the continuity of $\Psi$ and its gradient across the
perturbed boundary.  This is enforced by the following conditions (see
App.~\ref{app:interface_gravitational_potential}):
\begin{eqnarray}
\label{eq:interface_Psi}
\Psi_- &=& \Psi_+, \\
\label{eq:interface_grad_Psi}
\partial_{\zeta} \Psi_- + \frac{\Lambda \rho_- \zeta^2 \rz}{r^2+\rt^2} \xi^{\zeta} &=&
\partial_{\zeta} \Psi_+ + \frac{\Lambda \rho_+ \zeta^2 \rz}{r^2+\rt^2} \xi^{\zeta}.
\end{eqnarray}

\subsection{Boundary conditions}
\label{sect:boundary_conditions}

As usual, various boundary conditions are needed to complete the system.  In the
centre, the solutions need to be regular.  At the surface, we apply the simple
mechanical boundary condition $\delta p = 0$.  We note that in
\citet{Reese2013}, a more complex condition was imposed in order to have a
non-zero value for $\delta T/T_0$ at the surface, useful for mode visibility
calculations.  However, with such a condition, the pulsation equations do not
derive from a variational principle.  Here, since we are seeking to obtain
accurate frequencies, we prefer the simpler boundary condition ($\delta p = 0$),
so that we can then apply the variational principle as a supplementary check on
the accuracy.  Finally, the gravitational potential must match a vacuum
potential at infinity.  This is achieved by extending the gravitational
potential, thanks to Eqs.~(\ref{eq:interface_Psi})
and~(\ref{eq:interface_grad_Psi}), into an external domain which encompasses the
star and has a spherical outer boundary.  The outer boundary condition is then
\citep[see][]{Reese2006}:
\begin{equation}
\frac{1}{\rz} \dfrac{\Psi_m^{\l}}{\zeta} + \frac{\l+1}{r_{\mathrm{ext}}} \Psi_m^{\l} = 0,
\end{equation}
where $\rz = \dz r = 1 - \varepsilon$, $r_{\mathrm{ext}} = 2$, and where we have
used a harmonic decomposition of $\Psi$, $\l$ being the spherical harmonic
degree.

\subsection{Numerical approach}

The above system of equations, as well as boundary and interface conditions, are
discretised using the spherical harmonic basis for the angular coordinates
$(\theta, \phi)$, and using Chebyshev polynomials in the radial direction.  This
leads to a generalised matrix eigenvalue problem of the form $A x = \lambda B
x$. This problem is modified using a shift-invert approach to target frequencies
around a given shift, $\sigma$, before being solved through the
Arnoldi-Chebyshev approach \citep[\eg][]{Braconnier1993, Chatelin2012}.

The multi-domain spectral approach used in the radial direction leads to
matrices $A$ and $B$ which are block tri-diagonal.  The matrix $A-\sigma B$
(which intervenes in the shift-invert approach) can be efficiently factorised
using successive factorisations of the diagonal blocks (including a corrective
term from the non-diagonal blocks).

\subsection{Accuracy of the pulsation calculations}

\subsubsection{Various numerical resolutions}

In order to check the accuracy of the frequencies, it is useful to recalculate
the pulsation modes using different radial resolutions or numbers of spherical
harmonics.  Accordingly, we recalculated $28$ to $30$ axisymmetric ($m=0$) modes
in three of the models, using various resolutions. 
Table~\ref{tab:numerical_resolutions} gives the maximum relative differences on
the pulsation frequencies.  The first column corresponds to a $50\,\%$ increase
in the number of spherical harmonics in the pulsation calculations, that is, the
pulsation modes are calculated with $\Nt=60$ rather $\Nt=40$ spherical
harmonics.  The second column corresponds to a $\sim 50\,\%$ increase of the
radial resolution in the pulsation calculations (after having interpolated the
model). Specifically, the resolutions in the eight domains are 45, 85, 70, 60,
60, 75, 105, and 105, that is, a total of $\Nr=605$ points.  Finally, the third
column corresponds to $\sim 50\,\%$ increase of the radial resolution both in
the model (that is, the model is calculated with ESTER using an increased radial
resolution rather than being interpolated) and pulsation calculations.

\begin{table}[htbp]
\caption{Maximum relative differences on pulsation frequencies using various
resolutions in three of the models. \label{tab:numerical_resolutions}}
\begin{center}
\begin{tabular}{lccc}
\hline\hline
\textbf{Model} & 
$1.5 \times \Nt$ &
$1.5 \times \Nr$ &
$1.5 \times \Nrmod$ \\
\hline
\texttt{Mreal} & $1.2 \times 10^{-5}$ & $1.4 \times 10^{-8}$  & $3.2 \times 10^{-5}$ \\
\texttt{M}     & $1.2 \times 10^{-7}$ & $3.4 \times 10^{-12}$ & $7.6 \times 10^{-11}$ \\
\texttt{M6}    & $7.9 \times 10^{-4}$ & $2.8 \times 10^{-11}$ & --\\
\hline
\end{tabular}
\end{center}
\end{table}

Two trends can be seen in Table~\ref{tab:numerical_resolutions}.  First,
modifying the resolution in both the model and the pulsation calculations has a
greater impact than only modifying the resolution of the pulsation calculations.
This is expected as the higher resolution will be taken into account in the
ESTER convergence process when calculating the model in the former case. We note
that no value is provided in the last column of
Table~\ref{tab:numerical_resolutions} for model \texttt{M6} since ESTER was
unable to converge in that situation. Secondly, modifying the number of
spherical harmonics in the pulsation calculations has a greater impact than
modifying the radial resolution.  This probably simply illustrates the need for
a sufficient harmonic resolution to resolve the intricate island mode geometry,
particularly in model \texttt{M6}.  Overall, these differences remain small
(especially bearing in mind these are the maximal differences), except possibly
for the differences related to the harmonic resolution in model \texttt{M6}.

\subsubsection{Variational principle}
\label{sect:variational}

Another way of checking the accuracy of the pulsation calculations consists in
comparing the numerical frequencies with those obtained using a variational
formula.  Such a formula is an integral relation between the frequencies and
their associated eigenfunctions.  According to the variational principle, the
error on the variational frequency scales as the square of the error on the
eigenfunctions \citep[\eg][]{Christensen-Dalsgaard1982}. A general formulation
of the variational principle in differentially rotating bodies has previously
been obtained by \citet{Lynden-Bell1967}.  However, the formulation of some of
the terms, notably the use of Green's theorem for the gravitational potential,
is not the most suitable for numerical implementation.  Previous,
numerically-friendly expressions similar to those in \citet{Unno1989}, have been
obtained in \citet{Reese2006} and \citet{Reese2009a}, but these expressions were
only obtained for uniform or cylindrical rotation profiles, assumed that the
star is barotropic, and did not include the effects of discontinuities.  In
App.~\ref{app:variational_principle}, we give a full derivation for baroclinic
models with 2D rotation profiles and discontinuities. The final expression is:
\begin{eqnarray}
0 &=& \sum_i \mathop{\mathlarger{\int}}_{V_i} \mathop{\mathlarger{\mathlarger{\mathlarger{\mathlarger{\{}}}}} \vlp^2 \rho_0 \vect{\xi} \cdot \vect{\eta}^*
      - 2i\vlp \rho_0 \vect{\Omega} \cdot \left( \vect{\xi} \times \vect{\eta}^* \right) \nonumber \\
  & & -\rho_0 \left( \vect{\Omega} \cdot \vect{\xi} \right) \left( \vect{\Omega} \cdot \vect{\eta}^* \right) 
      + \rho_0 \Omega^2 \vect{\xi} \cdot \vect{\eta}^*
      - \vect{\eta}^* \cdot \left[ \vect{\xi} \cdot \grad \left( \grad P_0 \right) \right] \nonumber\\
  & & - \rho_0 \vect{\eta}^* \cdot \left[ \vect{\xi} \cdot \grad \left( \grad \Psi_0 \right) \right]
      - \frac{\pi^* P}{\Gamma_1 P_0}
      + \frac{\left(\vect{\xi} \cdot \grad P_0\right)\left(\vect{\eta}^* \cdot \grad P_0\right)}{\Gamma_1 P_0} \mathop{\mathlarger{\mathlarger{\mathlarger{\mathlarger{\}}}}}} \mathrm{dV} \nonumber \\
  & & + \sum_i \int_{S_i} \vect{\xi} \cdot \left(\grad P_0^- -\grad P_0^+\right) \vect{\eta}^* \cdot \vect{\mathrm{dS}}
      + \int_{V_{\infty}} \frac{\grad \Psi \cdot \grad \Phi^*}{\Lambda} \mathrm{dV},
\end{eqnarray}
where $\Lambda$ is $4\pi G$ or $4\pi$ in the dimensionless case, $V_i$ are the
different domains over which the stellar model is continuous, $S_i$ are the
surfaces of the discontinuities (including the stellar surface), the subscripts
`$+$' and `$-$' represent quantities right above and below the
discontinuities, respectively ($\grad P_0^+ = \vect{0}$ at the stellar surface), and
$V_{\infty}$ is infinite space (including the star).

\begin{figure}
\includegraphics[width=\columnwidth]{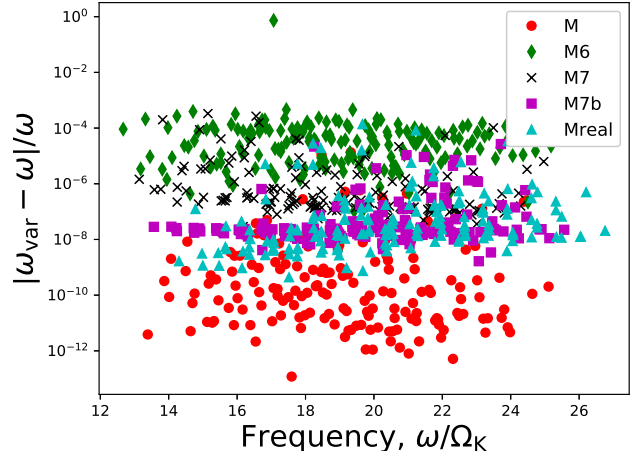}
\caption{Relative differences between numerical and variational frequencies.
\label{fig:variational}}
\end{figure}

Figure~\ref{fig:variational} shows the relative differences between the
numerical and variational frequencies for our models.  In each case a set of 168
modes with quantum numbers $\tilde{n}=19$ to $30$, $\tilde{\l}=0$ to $1$, $m=-3$
to $3$, was used.  We recall that $\tilde{n}$ is the number of nodes along an
island mode's orbit whereas $\tilde{\l}$ the number of nodes parallel to it.
These are related to the usual quantum numbers, $(n,\l,m)$, of pulsation modes
in the non-rotating case via the relations $\tilde{n} = 2 n + \varepsilon$ and
$\tilde{\l} = \frac{\l-|m|-\varepsilon}{2}$, where $\varepsilon \equiv
\l+m\,\mathrm{mod}\,2 \equiv \tilde{n}\,\mathrm{mod}\,2$ corresponds to the
mode's parity, that is, symmetry with respect to the equatorial plane
\citep{Reese2008b}. As can be seen in the figure, relative differences range
from $10^{-11}$ to $10^{-4}$, apart from an outlier in model
\texttt{M6}\footnote{We note that recalculating this mode with more spherical
harmonics brings the variational error to a level comparable with the other
modes.}. This compares quite favourably with the typical accuracy obtained with
space missions.  For instance, Kepler observations spanned up to four years
during the main mission thus leading to a Rayleigh resolution of
$0.008\,\mu$Hz.  The frequency at maximum amplitude of $\delta$ Scutis can reach
approximately $700\,\mu$Hz \citep[e.g.][]{Bowman2018} thus leading to a relative
precision as low as $10^{-5}$ in the best cases.  This is higher than the errors
on most of the variational frequencies, except for model \texttt{M6}.

The very high accuracy which is reached in a number of cases is due to the use
of spectral methods.  Such an accuracy was not reached straight away but rather
by repeating the calculation using the numerical frequency as the shift,
$\sigma$, in the second calculation and refining the solution through
supplementary iterations. Factors that decrease the accuracy (even in the second
calculations) are the presence of a discontinuity in the stellar model,
especially if it is sharp, and the occurrence of avoided crossings\footnote{We
recall that avoided crossings occur when the frequencies of two coupled modes
approach one another as a function of some stellar parameter such as age or
rotation rate.  Due to the coupling between the two modes, the frequencies do
not cross but the modes progressively exchange their geometric characteristics,
thereby leading to a mixture of the two geometries when the frequencies are
closest.  Figure~3 of \citet{Espinosa2004} provides a nice illustration of a
rotationally induced avoided crossing.} which lead to island modes which are
`polluted' by contributions from neighbouring modes.

\section{Rotation profile}
\label{sect:rotation}

\subsection{An approximate rotation kernel}

In order to understand the approximate effects of differential rotation on
pulsation frequencies, we consider a prograde acoustic mode and its retrograde
counterpart.  The azimuthal orders of these modes will be denoted $-m$ and $m$,
respectively\footnote{We are using the `retrograde' convention, that is,
retrograde modes have positive azimuthal orders.}.  Furthermore, the subscript
`+' will designate the prograde mode and `-' the retrograde mode.  The
variational principle can be expressed in the following approximate form for
these two modes:
\begin{equation}
0 \simeq (\omega_{\pm} \mp |m| \Omega_{\pm}^{\mathrm{eff}})^2 
      + 2(\omega_{\pm} \mp |m| \Omega_{\pm}^{\mathrm{eff}})\mathcal{C}_{\pm}
      + \mathrm{rest}_{\pm},
\label{eq:variational_schematic}
\end{equation}
where we have used the following definitions/approximations:
\begin{eqnarray}
\Omega_{\pm}^{\mathrm{eff}} &=& \frac{\int_V \Omega \rho_0 \|\vect{\xi}_{\pm}\|^2  dV}
                                     {\int_V \rho_0 \|\vect{\xi}_{\pm}\|^2 dV}, \\
\label{eq:Omega2eff_approx}
(\Omega_{\pm}^2)^{\mathrm{eff}} &=& \frac{\int_V \Omega^2 \rho_0 \|\vect{\xi}_{\pm}\|^2  dV}
                                     {\int_V \rho_0 \|\vect{\xi}_{\pm}\|^2 dV} 
                            \simeq \left(\Omega_{\pm}^{\mathrm{eff}}\right)^2, \\
\mathcal{C}_{\pm} &=& \frac{i\int_V \rho_0\vect{\Omega} \cdot\left(\vect{\xi}_{\pm}^* \times
\vect{\xi}_{\pm} \right) dV}
                           {\int_V \rho_0 \left\|\vect{\xi}_{\pm}\right\|^2 dV}, \\
\label{eq:Cor2_approx}
\left(\mathcal{C}_{\pm}\Omega_{\pm}\right)^{\mathrm{eff}} &=&
                   \frac{i\int_V \rho_0\Omega \vect{\Omega} \cdot\left(\vect{\xi}_{\pm}^* \times
\vect{\xi}_{\pm} \right) dV}
                           {\int_V \rho_0 \left\|\vect{\xi}_{\pm}\right\|^2 dV} 
                            \simeq \mathcal{C}_{\pm} \Omega_{\pm}^{\mathrm{eff}}.
\end{eqnarray}

If the two modes are of sufficiently high frequency so that the Coriolis force
only has a small impact, and if the rotation profile is not too differential,
then the two modes will be close to symmetric.  This means that by taking the
difference between Eq.~\ref{eq:variational_schematic} applied to the prograde
mode, and the same equation applied to the retrograde mode, the terms
`$\mathrm{rest}_+$' and `$\mathrm{rest}_-$' nearly cancel. Neglecting the
difference between these two terms leads to the following equation:
\begin{eqnarray}
(\omega_+ - |m| \Omega_+^{\mathrm{eff}})^2 
& + & 2(\omega_+ - |m| \Omega_+^{\mathrm{eff}})\mathcal{C}_+ \nonumber \\ & \simeq &
(\omega_- + |m| \Omega_-^{\mathrm{eff}})^2 
+ 2(\omega_- + |m| \Omega_-^{\mathrm{eff}})\mathcal{C}_-. 
\end{eqnarray}
This can be re-expressed as:
\begin{eqnarray}
(\omega_+ - |m| \Omega_+^{\mathrm{eff}})^2 &~&
\left[1 +  \frac{2\mathcal{C}_+}{(\omega_+ - |m| \Omega_+^{\mathrm{eff}})}\right]
\nonumber \\ & \simeq &
(\omega_- + |m| \Omega_-^{\mathrm{eff}})^2 
\left[1 + \frac{2\mathcal{C}_-}{(\omega_- + |m| \Omega_-^{\mathrm{eff}})}\right].
\end{eqnarray}
Taking the square-root of both sides and assuming $\mathcal{C}_{\pm} \ll (\omega_{\pm} \mp |m|\Omega_{\pm}^{\mathrm{eff}})$
leads to:
\begin{eqnarray}
(\omega_+ - |m| \Omega_+^{\mathrm{eff}}) &~&
\left[1 + \frac{\mathcal{C}_+}{(\omega_+ - |m| \Omega_+^{\mathrm{eff}})}\right] 
\nonumber \\ & \simeq &
(\omega_- + |m| \Omega_-^{\mathrm{eff}})
\left[1 + \frac{\mathcal{C}_-}{(\omega_- + |m| \Omega_-^{\mathrm{eff}})}\right].
\end{eqnarray}
This equation can finally be rearranged to yield:
\begin{equation}
\frac{\omega_+ - \omega_-}{2|m|} \simeq \frac{\Omega_+^{\mathrm{eff}} +
\Omega_-^{\mathrm{eff}}}{2} + \frac{-\mathcal{C}_+ + \mathcal{C}_-}{2|m|}.
\label{eq:OmegaEff}
\end{equation}
This equation is particularly interesting because it provides a linear relation
between the generalised rotational splitting, which only depends on the
frequency of the modes, and the rotation profile.  The weighting function that
intervenes in the integral is known as the rotation kernel and only depends on
the eigenfunctions. If the Coriolis force is neglected, this equation reduces to
the linearised version of Eq.~(32) from \citet{Reese2009a}.

\begin{figure}[htbp]
\includegraphics[width=\columnwidth]{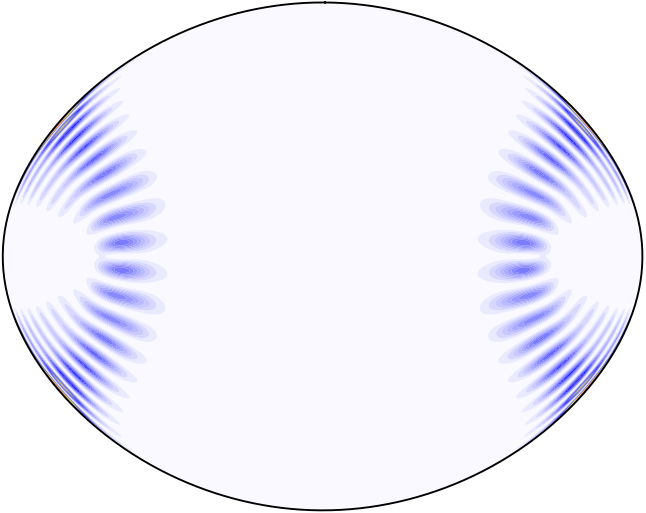}
\caption{Rotation kernel for an island mode in model \texttt{Mreal}.
The island mode is an $m=1$ mode which is symmetric with respect
to the equator and has a frequency of $\nu = 760.6\,\mu\mathrm{Hz}$.
\label{fig:rotation_kernel}}
\end{figure}

Figure~\ref{fig:rotation_kernel} shows what a typical rotation kernel will look
like for an island mode.  As can be seen, the rotation kernel closely follows
the geometry of the island mode much like in \citet{Reese2009a}.  Accordingly,
these modes are especially sensitive to the rotation rate in this region, in
particular near the surface at mid-latitudes.

\begin{figure}[htbp]
\includegraphics[width=\columnwidth]{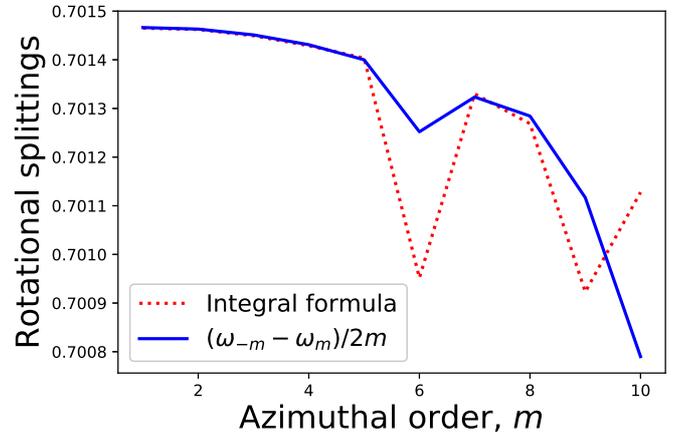}
\caption{Comparison between generalised rotational splittings and the corresponding
weighted integrals of the rotation profile for model \texttt{M} (see
Eq.~\ref{eq:OmegaEff}).\label{fig:splittings}}
\end{figure}

\begin{figure*}[htbp]
\begin{center}
\includegraphics[width=0.49\textwidth]{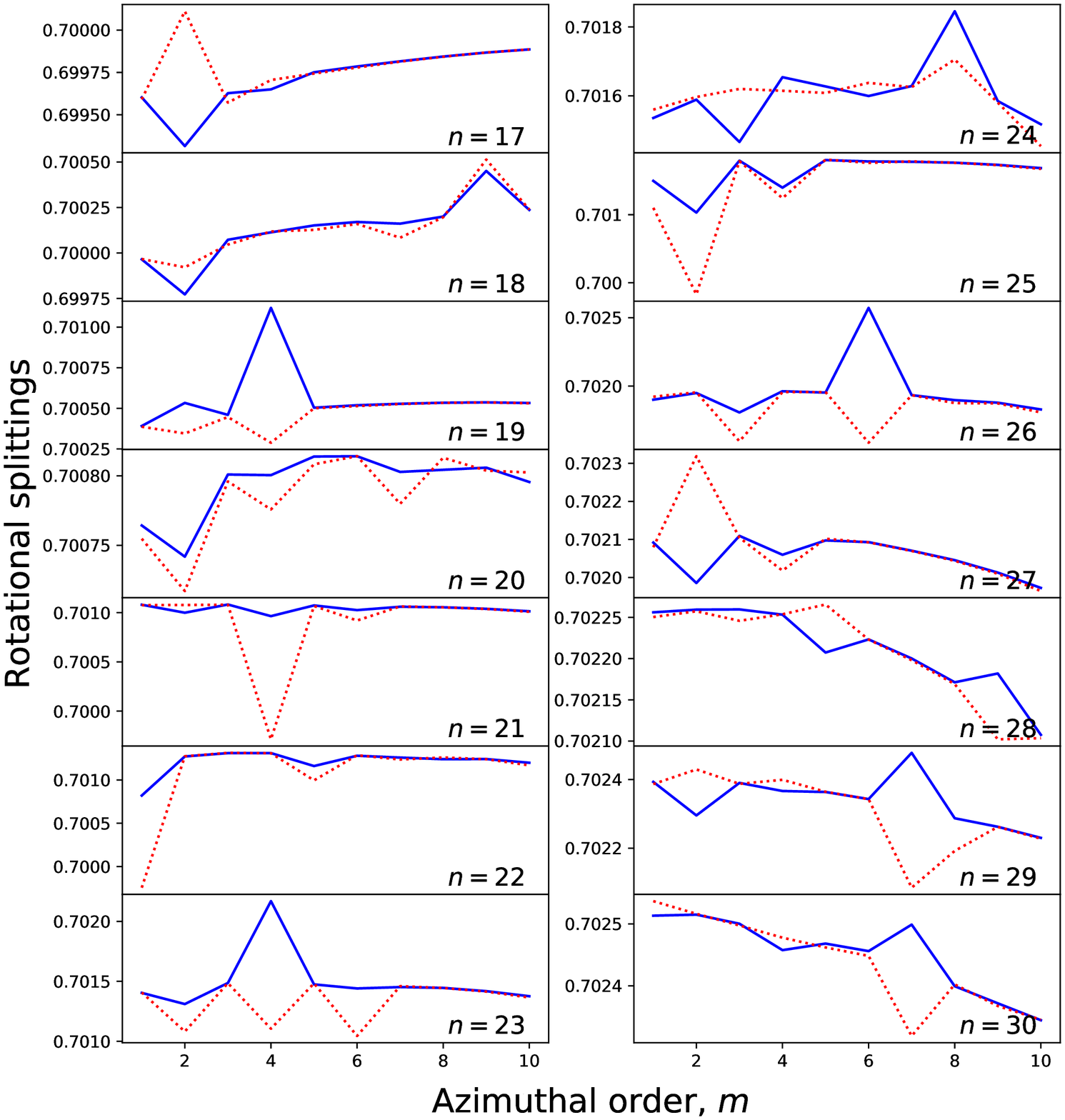} 
\includegraphics[width=0.49\textwidth]{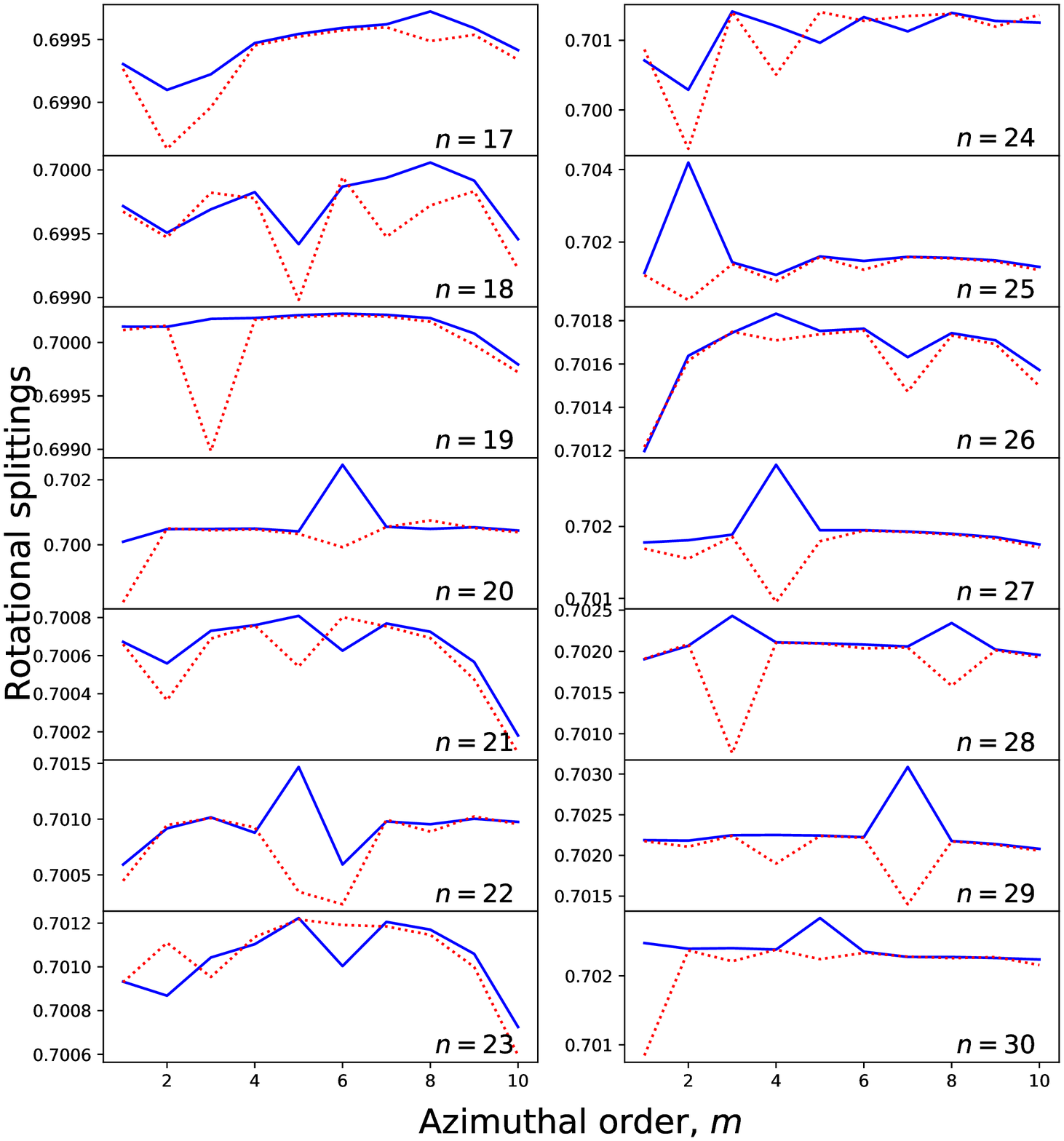}
\end{center}
\caption{Same as Fig.~\ref{fig:splittings} but for model \texttt{Mreal}
and a more extensive set of modes.  \label{fig:splittings_bis}}
\end{figure*}

\begin{figure*}[htbp]
\begin{tabular}{cc}
\textbf{Prograde} &
\textbf{Retrograde} \\
\includegraphics[width=0.48\textwidth]{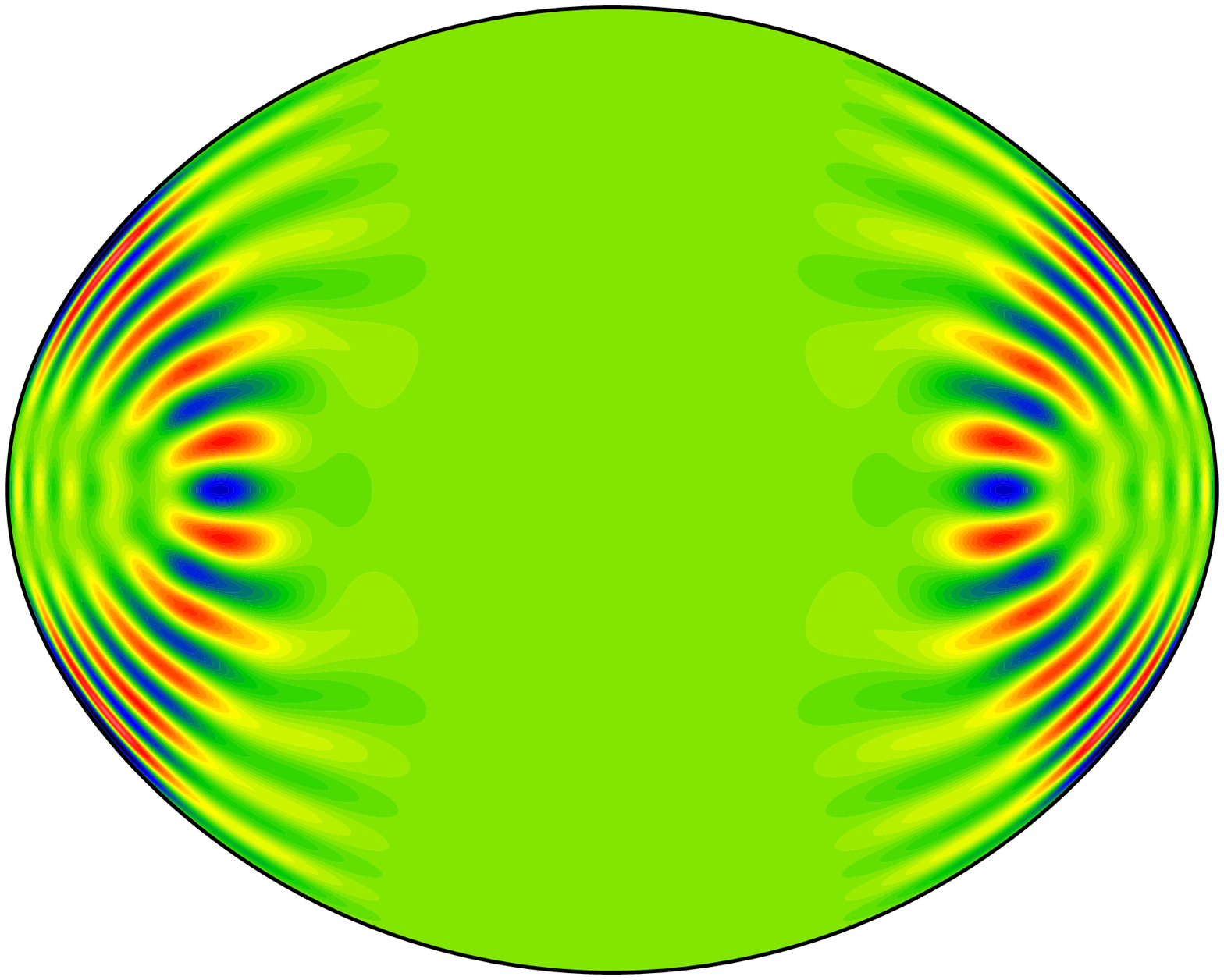} &
\includegraphics[width=0.48\textwidth]{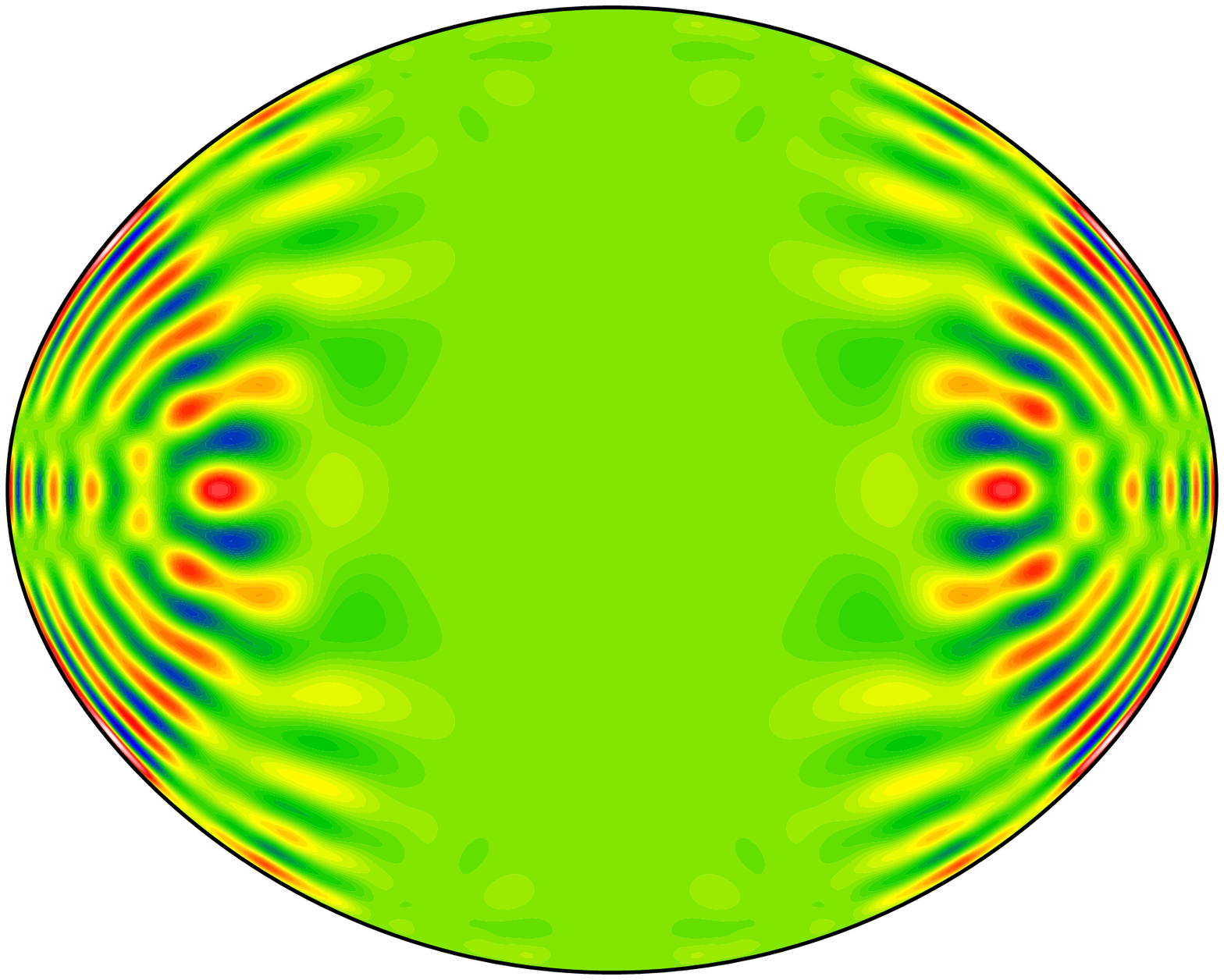}
\end{tabular}
\caption{Meridional cross-sections of a prograde mode with $m=-6$ and its
retrograde counterpart.  The frequencies of these modes are, respectively,
$610.2$ and $358.8$ $\mu$Hz. These modes are involved in avoided crossings with
other modes (not shown). As a result of being at different stages of the
avoided crossing, their geometry is different and applying
Eq.~(\ref{eq:OmegaEff}) yields less accurate results.
\label{fig:splittings_avoided_crossing}}
\end{figure*}

In Figs.~\ref{fig:splittings} and~\ref{fig:splittings_bis}, we compare the
generalised splittings with the right-hand sides of Eq.~(\ref{eq:OmegaEff}) for
models \texttt{M} and \texttt{Mreal}, respectively.  The latter is for a much
more extensive set of modes.  As can be seen, a good agreement is obtained in
most cases, but there are some notable exceptions.  Such exceptions typically
occur for avoided crossings.  Indeed, the geometry of the modes changes rapidly
as a function of the rotation rate during avoided crossings thereby causing
prograde modes and their retrograde counterparts to be at different parts of
their avoided crossings and to have different geometric structures. 
Figure~\ref{fig:splittings_avoided_crossing} provides an example of such modes. 
As a result, the terms `$\mathrm{rest}_+$' and `$\mathrm{rest}_-$' do not cancel
each other out.  This interpretation is confirmed in Table~\ref{tab:splittings}
which provides a detailed comparison between modes in this situation (Solutions
3 and 4) and those which are not undergoing an avoided crossing (Solutions 1 and
2).  By including the difference between the terms `$\mathrm{rest}_+$' and
`$\mathrm{rest}_-$' , it is possible to correct Eq.~(\ref{eq:OmegaEff}) and
improve the agreement by a factor of 20 for Solutions 3 and 4.  The
supplementary rows in this Table also show that the approximations given in
Eqs.~(\ref{eq:Omega2eff_approx}) and~(\ref{eq:Cor2_approx}) are well justified.
Hence, apart from the cases involving avoided crossings, the agreement between
the generalised splittings and the weighted integrals of the rotation profile
(that is, the right-hand side of Eq.~(\ref{eq:OmegaEff})) is excellent thus
potentially providing the basis for probing the rotation profile via inversions.


\begin{table*}[htbp]
\caption{Generalised splittings versus weighted integrals of the rotation
profile, and different terms from Eq.~(\ref{eq:variational_schematic}) for two
pairs of prograde and retrograde modes.  $\delta\omega_{\mathrm{var}}/\omega$
corresponds to the relative error on the variational frequency, and
$\delta\omega_{\mathrm{var}}^{\mathrm{approx.}}/\omega$ is the same error when
$\omega_{\mathrm{var}}$ is calculated using the approximations in
Eqs.~(\ref{eq:Omega2eff_approx}) and~(\ref{eq:Cor2_approx}). The modes from the
second pair, Solutions~3 and~4, are undergoing avoided crossings, whereas the
other two modes are not. \label{tab:splittings}}
\begin{center}
\begin{tabular}{lcccc}
\hline
\hline
\textbf{Quantity}& \textbf{Solution 1}& \textbf{Solution 2}& \textbf{Solution 3}& \textbf{Solution 4}\\
\hline
$m$& $-2$ & $2$ & -6 & 6 \\
$\omega$& 17.50030 & 14.69444 & 20.42777 & 12.01274 \\
$\frac{\omega_+ - \omega_-}{2|m|}$ & \multicolumn{2}{c}{0.70146} & \multicolumn{2}{c}{0.70125} \\
$\frac{\Omega_+^{\mathrm{eff}} + \Omega_-^{\mathrm{eff}}}{2} + \frac{-\mathcal{C}_+ + \mathcal{C}_-}{2|m|}$ & \multicolumn{2}{c}{0.70146} & \multicolumn{2}{c}{0.70095}\\
\hline
$\delta\omega_{\mathrm{var}}/\omega$& $-8.50 \times 10^{-11}$ & $-5.18 \times 10^{-11}$ & $-3.42 \times 10^{-10}$ & $-1.09 \times 10^{-6}$ \\
$\delta\omega_{\mathrm{var}}^{\mathrm{approx.}}/\omega$ & $2.01 \times 10^{-7}$ & $-9.84 \times 10^{-8}$ & $5.34 \times 10^{-7}$ & $-4.42 \times 10^{-8}$ \\
$\Omega_{\pm}^{\mathrm{eff}}$& 0.70463 & 0.70464 & 0.70438 & 0.70436 \\
$\left(\Omega_{\pm}^2\right)^{\mathrm{eff}}$ & 0.49651 & 0.49652 & 0.49615 & 0.49614 \\
$\left(\Omega_{\pm}^{\mathrm{eff}}\right)^2$ & 0.49651 & 0.49651 & 0.49615 & 0.49613 \\
$\mathcal{C}_{\pm}$ & -0.04586 & -0.05856 & -0.03770 & -0.07873 \\
$\left(\mathcal{C}_{\pm}\Omega_{\pm}\right)^{\mathrm{eff}}$ & -0.03234 & -0.04128 & -0.02657 & -0.05544 \\
$\mathcal{C}_{\pm}\Omega_{\pm}^{\mathrm{eff}}$ & -0.03232 & -0.04126 & -0.02656 & -0.05545\\
rest$_{\pm}$ & -257.94181 & -257.94021 & -261.76369 & -261.64204 \\
\hline
\end{tabular}
\end{center}
\end{table*}

\section{Acoustic glitches}
\label{sect:glitches}

We now turn our attention to pulsations in the discontinuous models and focus on
acoustic glitches. We recall that glitches are regions in the star with a strong
gradient or near discontinuity, which can lead to an oscillatory behaviour in
the pulsation spectrum \citep[\eg][]{Monteiro1994}.

\subsection{Frequencies}

\begin{figure*}
\includegraphics[width=\textwidth]{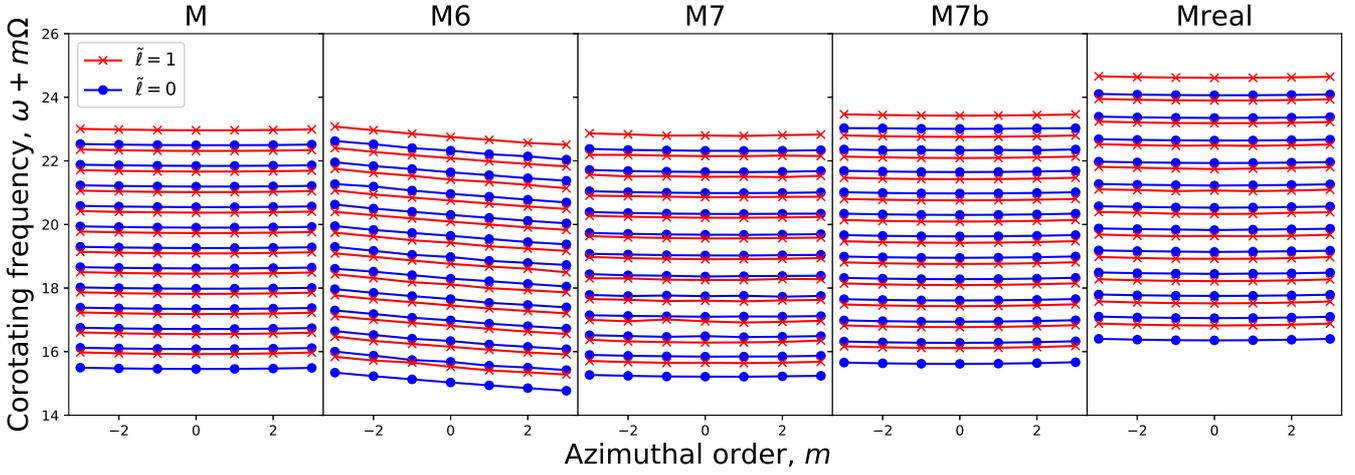}
\caption{Corotating pulsation frequencies in models \texttt{M},
\texttt{M6},\texttt{M7},\texttt{M7b}, and \texttt{Mreal}.
These frequencies are fairly well described by the empirical formula
from \citet{Reese2009a}. \label{fig:frequencies_glitches}}
\end{figure*}

\begin{figure}
\includegraphics[width=\columnwidth]{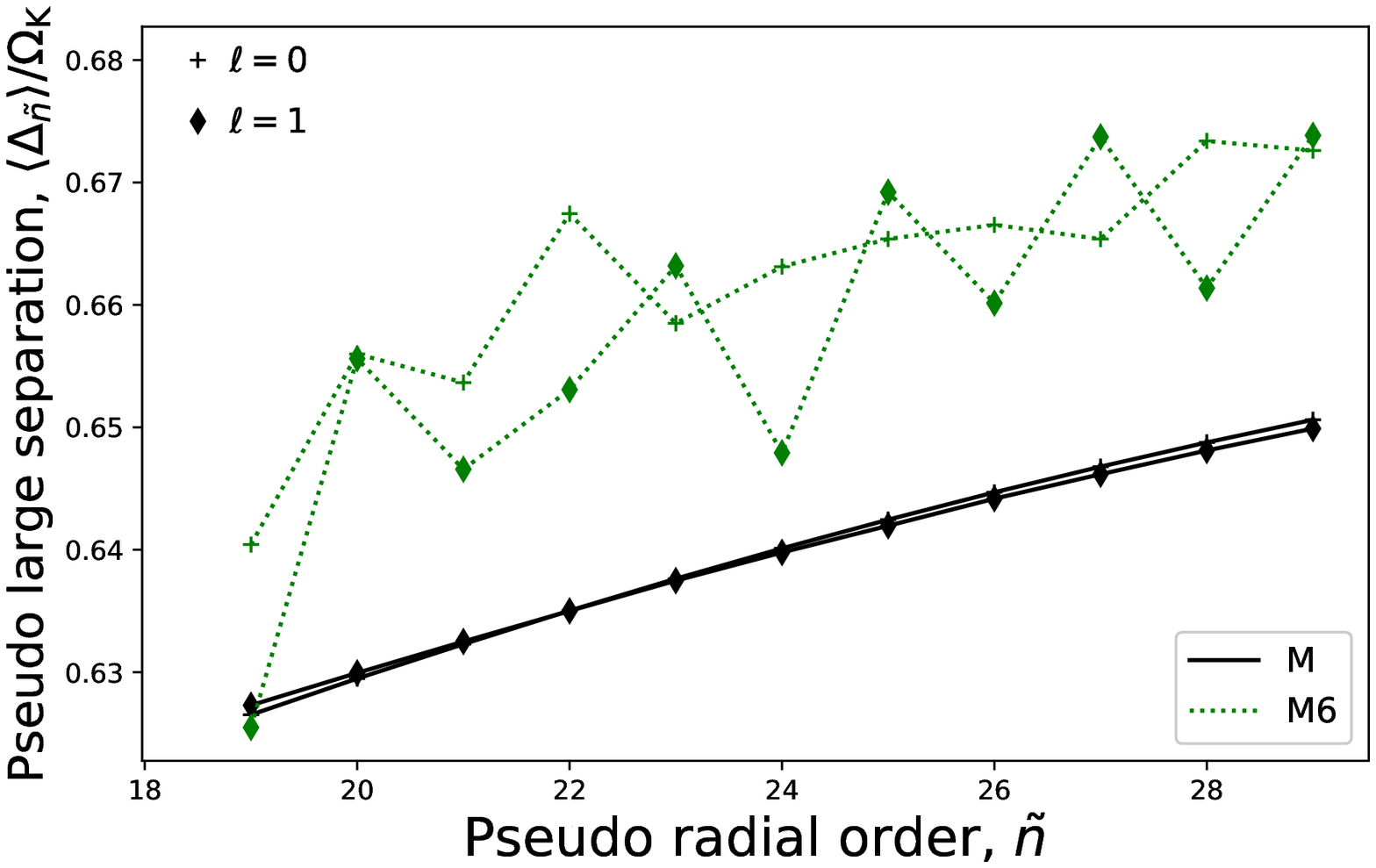}
\includegraphics[width=\columnwidth]{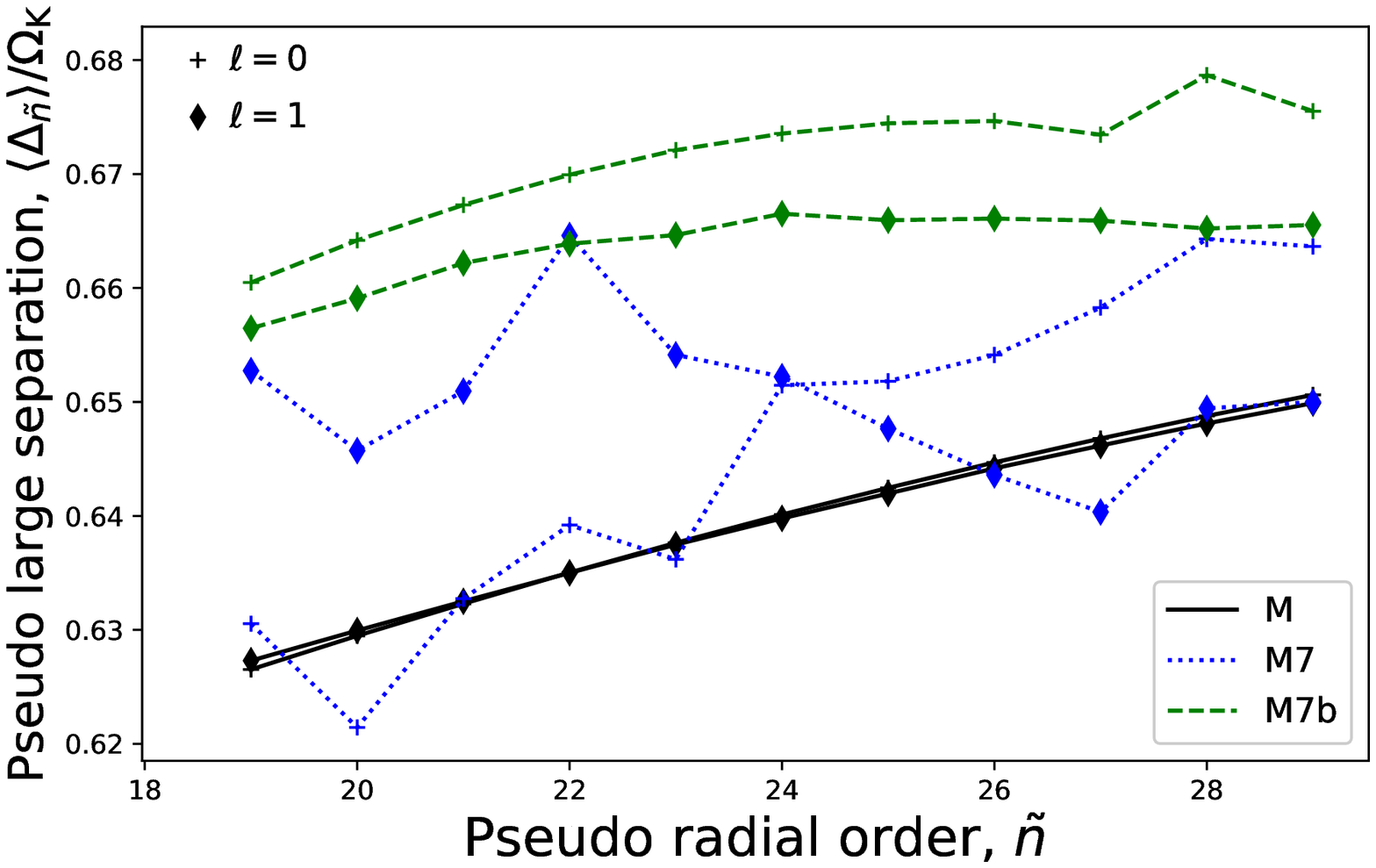}
\caption{Averaged pseudo large frequency separations,
$\left<{\Delta}_{\tilde{n}}\right>$, for the different models.
\label{fig:delta_n}}
\end{figure}

Figure~\ref{fig:frequencies_glitches} shows the pulsation frequencies obtained
for the various models for modes with $\tilde{n} = 19$ to $30$, $\tilde{\l}=0$
to $1$, and $m=-3$ to $3$.  As can be seen, these frequencies follow fairly
closely the asymptotic formula given in \citet{Reese2009a}.  However, a closer
look reveals irregularities in the pulsation spectra of the discontinuous
models.  This is brought out more clearly with the frequency separations
$\Delta_{\tilde{n}} = \omega_{\tilde{n}+1,\,\tilde{\l},\,m} -
\omega_{\tilde{n},\,\tilde{\l},\,m}$.  In Fig.~\ref{fig:delta_n}, we plot
averaged large separations, $\left<\Delta_{\tilde{n}}\right> =
\left<\omega_{\tilde{n}+1,\,\tilde{\l}}\right> - 
\left<\omega_{\tilde{n},\,\tilde{\l}}\right>$, where
$\left<\omega_{\tilde{n},\,\tilde{\l}}\right>$ is the pulsation frequency
averaged over the azimuthal orders $m=-3$ to $3$.  This is done in order to
reduce the effects of avoided crossings which tend to be more numerous in the
discontinuous models and tend to mask the frequency variations caused by the
glitch.  Even then, the averaged large separations in the discontinuous models
are more irregular than in the continuous model.  This raises the question
whether these variations can be explained by glitch theory.

\subsection{Glitch analysis and ray dynamics}

\begin{figure}[htbp]
\includegraphics[width=\columnwidth]{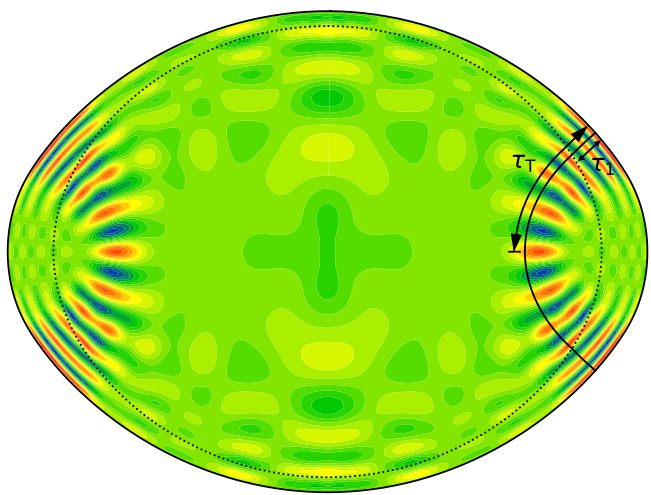}
\caption{Island mode with periodic orbit superimposed, in model \texttt{M6}.
The mode is axisymmetric with a frequency of 309.2 $\mu$Hz.
The dotted line shows the location of the discontinuity.  The acoustic travel
times $\tauT$ and $\tau_1$ along the ray trajectory are illustrated.
\label{fig:island_orbit}}
\end{figure}

In order to investigate the behaviour of the frequencies in a more detailed
way, we carried out a simplified ray dynamics analysis.  We used the following
dispersion relation, valid for axisymmetric modes in the high frequency limit:
\begin{equation}
\omega^2 = c_0^2 k^2,
\end{equation}
where $k$ is the norm of the wave-vector. A simple reflection was used at the
stellar surface, rather than a more realistic but complex approach involving the
cut-off frequency \citep[\eg][]{Lignieres2009}.  Furthermore, we applied the
Snell-Descartes refraction law at the discontinuity:
\begin{equation}
\frac{\sin\vartheta_+}{c_+} = \frac{\sin\vartheta_-}{c_-},
\end{equation}
where $\vartheta$ is the angle between the surface normal and the wave-vector,
$c$ the local sound velocity, and the subscripts `$+$' and `$-$' the upper
and lower domains at the discontinuity.  We neglect the partial wave reflection
at the discontinuity, since we are only searching for the island mode periodic
orbit.  A more complete description of the ray dynamics is provided in
App.~\ref{app:ray_dynamics}.  Figure~\ref{fig:island_orbit} shows the periodic
orbit for island modes superimposed on an island mode in model \texttt{M6}. 
As can be seen, the orbit reproduces very well the location of the mode.

\begin{figure*}[htbp]
\includegraphics[width=\textwidth]{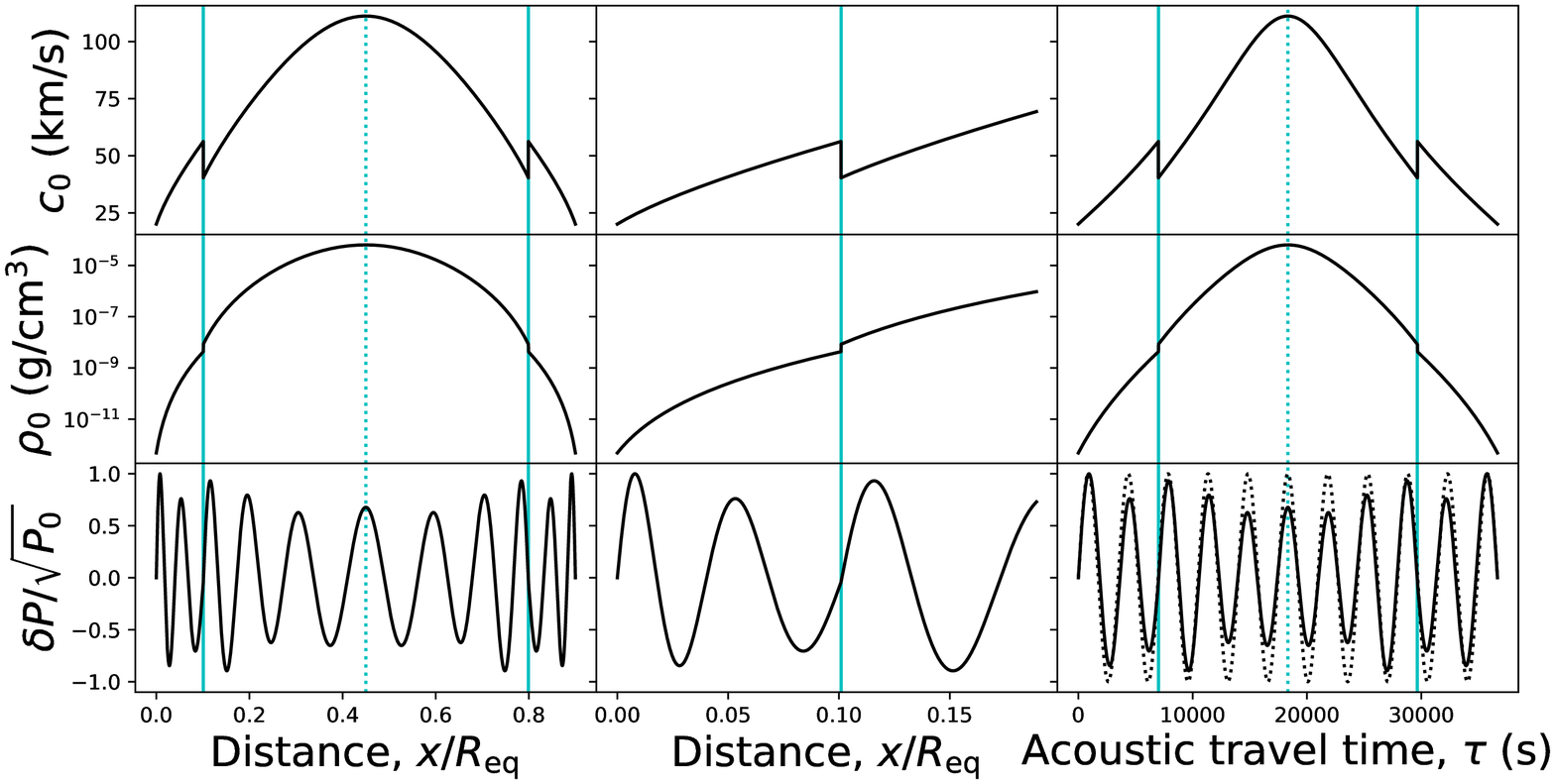}
\caption{Sound velocity (first row), density (second row), and perturbed
pressure ($\delta p/\sqrt{P_0}$, third row) profiles as a function of distance
(first and second columns) and acoustic travel time (third column) along island
periodic orbit. The second column is a zoom of the first column around the first
discontinuity. The vertical light blue solid lines indicate the discontinuity
and the vertical light blue dotted lines correspond to the equator.  The dotted
curve in the lower right panel is a simple sine curve with the same periodicity
as the mode. \label{fig:profiles}}
\end{figure*}

Figure~\ref{fig:profiles} then shows the sound velocity, density, and perturbed
pressure ($\delta p/\sqrt{P_0}$) profiles calculated along the periodic orbit,
both as a function of distance along the profile and acoustic travel time.  As
expected, a sharp transition in wavelength occurs at the discontinuity.
Furthermore, when plotted as a function of acoustic travel time, the wave takes
on a nearly sinusoidal behaviour as indicated by the comparison with the simple
sine curve, apart from a phase shift at the discontinuity and a variable
amplitude.

These observations provide the basis for a simple toy model which is described
in App.~\ref{app:toy_model}.  According to this model, the frequencies are given
to first order by:
\begin{equation}
\label{eq:toy_model}
\omega = \frac{1}{2\tauT} \left[ n\pi + \epsilon \sin\left(n\pi\frac{\tau_1}{\tauT}\right) \right],
\end{equation}
where $\tauT=\int_{\mathrm{surf.}}^{\mathrm{eq.}} \frac{dr}{c}$ is the acoustic
travel time from the surface to the equator along the ray path, and
$\tau_1=\int_{\mathrm{surf.}}^{\mathrm{disc.}} \frac{dr}{c}$, the acoustic
travel time from the surface to the discontinuity, as illustrated in
Fig.~\ref{fig:island_orbit}.  The quantity $\epsilon$ is given by the relation:
\begin{equation}
\epsilon = \frac{k_-}{k_+} - 1 = \frac{c_+}{c_-} - 1
\end{equation}
and is treated as a small parameter.  As shown in App.~\ref{app:toy_model}, even
for $\epsilon = 0.39$ (for model \texttt{M7}), Eq.~(\ref{eq:toy_model}) gives an
accurate estimate of the glitch period and a rough idea of its amplitude.
However, we do not expect the toy model to give an accurate idea of the
phase of the glitch pattern on the oscillation frequencies as it would require
fully treating surface effects.

\begin{table}[htbp]
\begin{center}
\caption{Acoustic travel times in various models. The quantities
$\tauT$ and $\tau_1$ are illustrated in Fig.~\ref{fig:island_orbit}.
\label{tab:travel_times}}
\begin{tabular}{ccc}
\hline
\hline
\textbf{Model name} &
$\tauT$ (in s) &
$\tau_1$ (in s) \\
\hline
\texttt{M}     & 12240.8 & -- \\
\texttt{M6}    & 18330.7 & 7003.8 \\
\texttt{M7}    & 16538.4 & 2153.3 \\
\texttt{M7b}   & 12676.9 &  787.9 \\
\texttt{Mreal} &  6612.9 &  670.4\tablefootmark{\dag} \\
\hline
\end{tabular}
\tablefoot{\tablefoottext{\dag} This corresponds to the He II
ionisation zone rather than to a discontinuity.}
\end{center}
\end{table}

\begin{figure}[htbp]
\includegraphics[width=\columnwidth]{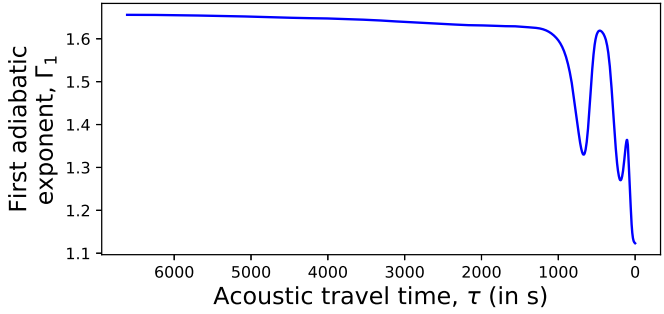}
\caption{$\Gamma_1$ profile in model \texttt{Mreal} along the island
mode orbit.  The stellar surface corresponds to $\tau=0$ s on the right side,
and the intersection of the equatorial plane with the orbit to $\tau=6612.9$ s
on the left side. \label{fig:Gamma1_realistic}}
\end{figure}

\begin{figure}[htbp]
\includegraphics[width=\columnwidth]{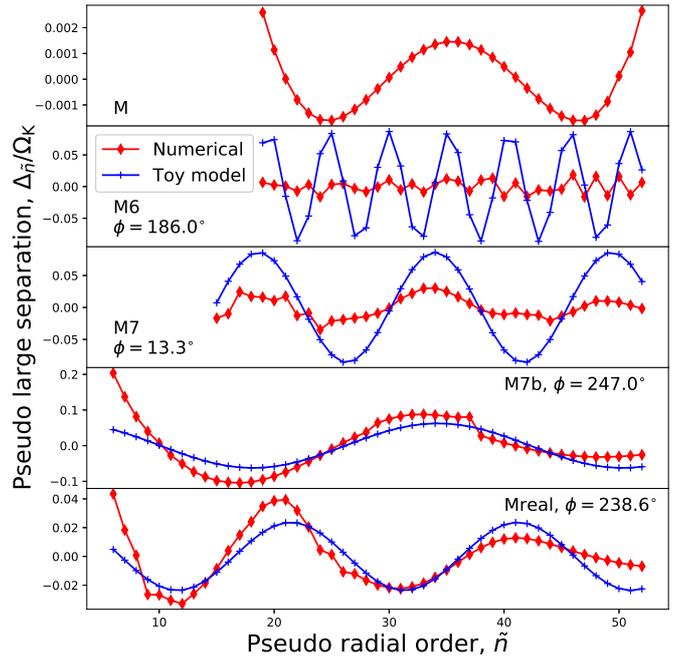}
\caption{Frequencies of the $(\tilde{\l},m) = (0,0)$ modes after
subtraction of a third-order polynomial fit (or second order polynomial fit in
the case \texttt{M7b}) versus the predicted glitch from the toy model. Each
panel corresponds to a different model.  An ad hoc phase, the value of which is
indicated in each panel, has been added to the toy model to improve the
agreement.\label{fig:delta_vs_toy}}
\end{figure}

Table~\ref{tab:travel_times} provides the acoustic travel times $\tauT$ and
$\tau_1$ for the different models in our study.  Although model
\texttt{Mreal} is continuous, we included the $\tau_1$ value for the He II
ionisation zone. Indeed, the $\Gamma_1$ profile undergoes a dip in that region,
as illustrated in Fig.~\ref{fig:Gamma1_realistic}.  Based on these values,
Fig.~\ref{fig:delta_vs_toy} compares the predictions from the toy model with the
$(\tilde{l},m) = (0,0)$ frequencies minus a second or third-order polynomial fit
in order to isolate the glitch pattern.  Indeed, using the large separations
rather than the frequencies would tend to amplify the impact of avoided
crossings thus making it harder to see the glitch pattern.  We note that a
second rather than third-order polynomial fit was used for model \texttt{M7b}
given the relatively long period of the glitch pattern which can be mimicked up
to some extent by a third- or higher-order polynomials.  An ad hoc phase was
added to the glitch pattern from the toy model given that this model is not
expected to correctly predict the phase as described above.  This allows us to
focus on the period and amplitude of the glitch pattern to see how accurate the
predictions are.

As can be seen from Fig.~\ref{fig:delta_vs_toy}, a nice agreement is
obtained for models \texttt{M7b}, \texttt{Mreal}, and to a lesser extent
\texttt{M7}. This confirms that the toy model is able to correctly predict the
periodicity of the glitch pattern, at least in some cases.  The agreement on the
amplitude is satisfactory for \texttt{M7b} but rather poor for \texttt{M7}.  For
model \texttt{Mreal}, an ad hoc amplitude was used for the predicted glitch
pattern.  Indeed, the toy model was specifically constructed for discontinuities
and is therefore unable to predict the amplitude of the glitch pattern for a
smoother transition such as what takes place in an ionisation zone.  It is
nonetheless interesting to note that the amplitude of this glitch pattern
decreases at higher frequencies, as would be expected for such a transition.
Model \texttt{M} is not expected to show a glitch pattern since it contains no
discontinuities and the $\Gamma_1$ profile is very close to $5/3$ throughout the
star, as a result of the ideal gas equation of state.  The plot shows what is
likely to be a fourth order polynomial residual as expected when subtracting a
third-order polynomial fit, as confirmed by the much smaller scale of the
$y$-axis. In contrast, no agreement is found between the toy model and the
glitch pattern for model \texttt{M6}. The reasons for this lack of agreement are
not entirely understood, but we do note that most of its island modes are
undergoing avoided crossings in contrast to the other models.  Avoided crossings
typically cause the frequencies to deviate from their asymptotic values and
could therefore easily mask a glitch pattern.

\subsection{Pulsation mode geometry at the discontinuity}

We now investigate in a detailed way the local geometric properties of
the islands modes in the region where the periodic orbit intersects the
discontinuity.  Specifically, we check whether the wave amplitudes match the
predictions from a local analysis, and whether the angle between the
discontinuity and the orbit matches a numerical estimate based on the island
mode.

As recalled in App.~\ref{app:refraction}, the pulsation mode including the 
reflected and refracted waves can locally be approximated as:
\begin{equation}
\left(\dP\right)^{\pm} = \left[A_1^{\pm} \cos(\vect{k}_1^{\pm} \cdot \vect{x})
                       + A_2^{\pm} \cos(\vect{k}_2^{\pm} \cdot \vect{x}) \right]
                       \exp(i\omega t),
\end{equation}
where the superscripts `$+$' and `$-$' designate the upper and lower
domains, respectively, $\vect{k}_1^{\pm}$ and $\vect{k}_2^{\pm}$ wave vectors,
and where the amplitudes, $A_1^{\pm}$, $A_2^{\pm}$, are
related via the relation:
\begin{equation}
\label{eq:amplitudes}
\left[
\begin{array}{c}
A_1^+ \\
A_2^+
\end{array}
\right]
=
\frac{1}{2}
\left[
\begin{array}{cc}
1 + \eta & 1 - \eta \\
1 - \eta & 1 + \eta
\end{array}
\right]
\left[
\begin{array}{c}
A_1^- \\
A_2^-
\end{array}
\right],
\end{equation}
where
\begin{equation}
\eta = \frac{\rho_0^+}{\rho_0^-} \frac{\kperp^-}{\kperp^+}.
\end{equation}
and $\kperp^{\pm}$ is the wave vector component perpendicular to the
surface.  When $\kpar \ll \kperp$, the tangential component, the factor
$\eta$ reduces to:
\begin{equation}
\eta \simeq \sqrt{\frac{\rho_0^+}{\rho_0^-}} = \frac{c_0^-}{c_0^+}.
\end{equation}

\begin{figure*}[htbp]
\begin{minipage}{0.48\textwidth}
\includegraphics[width=\columnwidth]{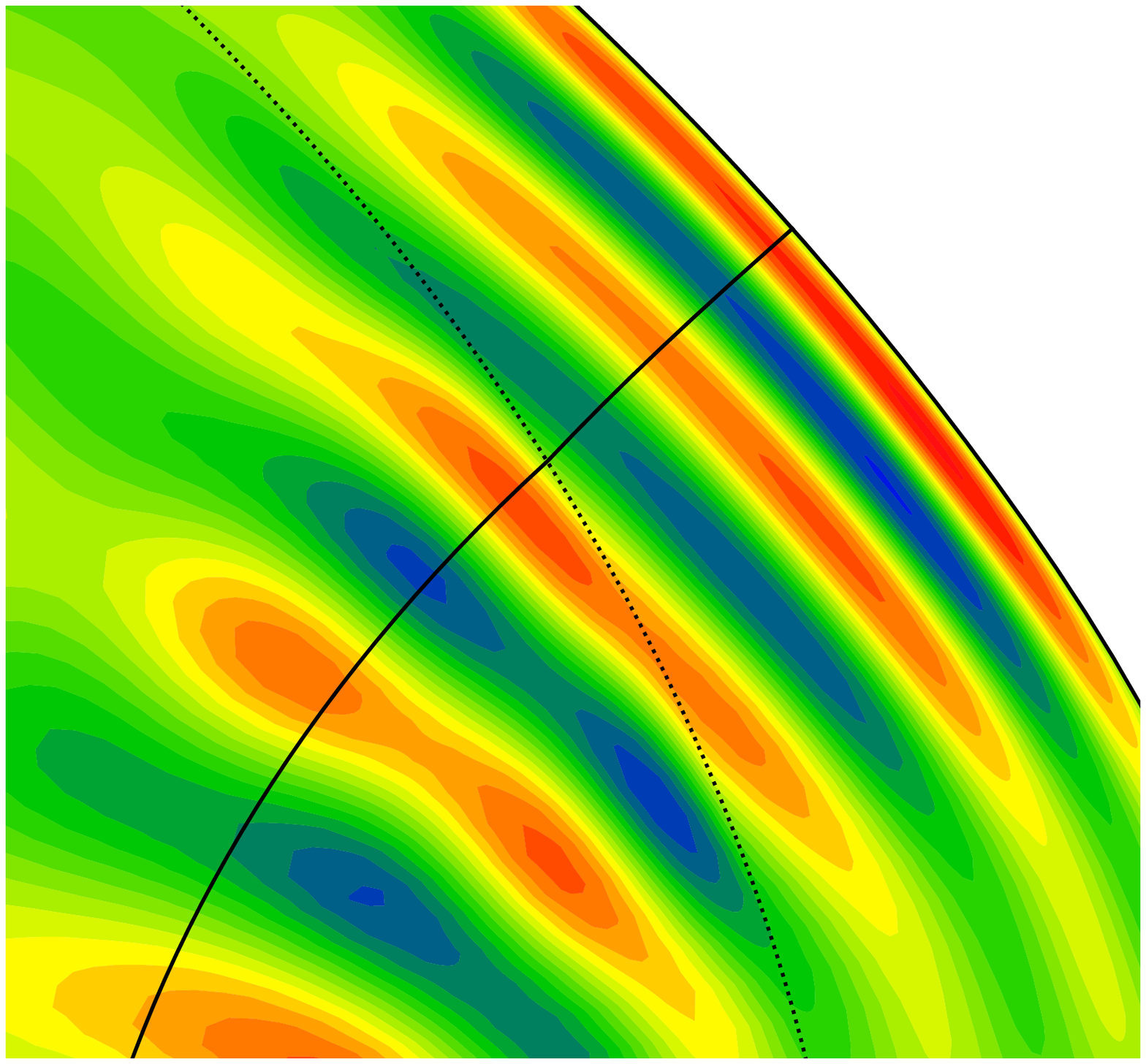}
\caption{Zoom in on the island mode shown in Fig.~\ref{fig:island_orbit} (in
model \texttt{M6}).  The discontinuity is shown using the dotted line, and
the solid line corresponds to the island mode orbit. As can be seen, the
wavelength decreases just above the discontinuity. \label{fig:field_patch}}
\end{minipage} \hfill
\begin{minipage}{0.48\textwidth}
\includegraphics[width=\columnwidth]{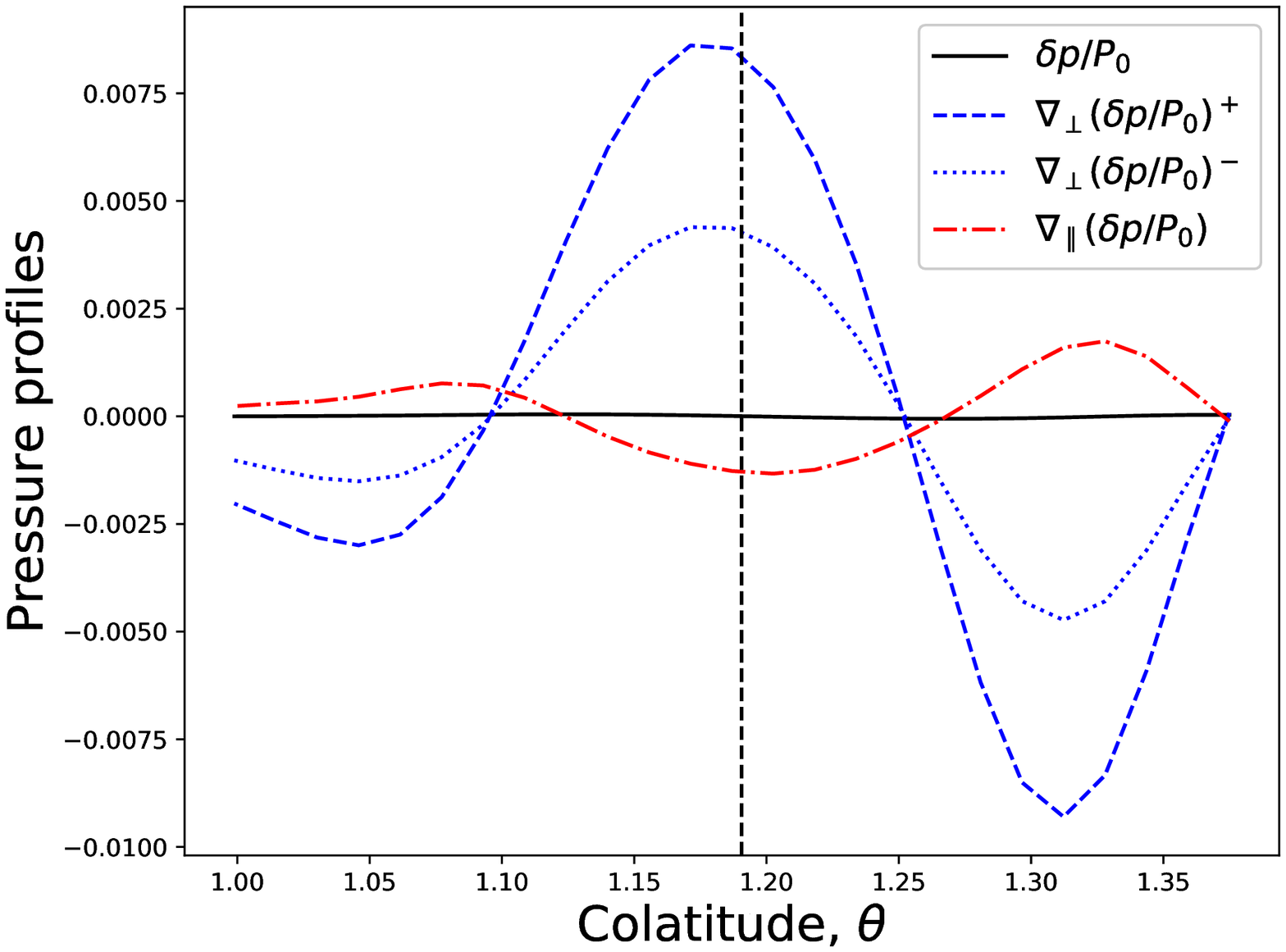}
\caption{Extracted $\dPi$, $\grad_{\parallel} (\dPi)$, and $\grad_{\perp}
(\dPi)^{\pm}$ profiles as a function of $\theta$ along the discontinuity shown
in Fig.~\ref{fig:field_patch}.  The vertical dashed line shows colatitude where
the island mode periodic orbit crosses the discontinuity.
\label{fig:extracted_profiles}}
\end{minipage}
\end{figure*}

We then investigate several $m=0$ island modes in different models to extract
the amplitudes of the refracted and reflected waves and verify the above
equations.  We start by extracting the $\dPi$ profile as well as its horizontal
and vertical gradients, $\grad_{\parallel} (\dPi)$, and $\left(\grad_{\perp}
(\dPi)\right)^{\pm}$, just above and below the discontinuity\footnote{We recall
that $\left(\grad_{\parallel} (\dPi)\right)^+=\left(\grad_{\parallel}
(\dPi)\right)^-$ as a result of the continuity of $\dPi$.}.  Since we are
focusing on axisymmetric modes, the horizontal gradient is in the meridional
plane -- there is no component in the $\vect{e}_{\phi}$ direction.
Figure~\ref{fig:field_patch} shows a zoom on part of an island mode in model
\texttt{M6} and Fig.~\ref{fig:extracted_profiles} shows the extracted
profiles.  The amplitudes of these profiles are estimated thanks to their
maximum absolute values.  Given that the $\left(\grad_{\perp}
(\dPi)\right)^{\pm}$ profiles have the opposite sign to the $\grad_{\parallel}
(\dPi)$ profile, these have negative amplitudes.  The tangential wave vector
component (which is the same above and below) is estimated by calculating the
ratio between the amplitudes of $\grad_{\parallel} (\dPi)$ and $\dPi$.  The
normal components above and below the discontinuity are obtained via the
dispersion relation, thus enforcing Snell-Descartes' law. The individual wave
amplitudes are obtained by calculating appropriate linear combinations of
$\left(\grad_{\perp} (\dPi)\right)^{\pm}/\kperp$ and $\left(\grad_{\parallel}
(\dPi)\right)/\kpar$.

Table~\ref{tab:wave_parameters} gives the wave vector components and amplitudes
for island modes in three of the models.  Given that the mode amplitude is
arbitrary, we normalised the amplitudes by $A_2^-$.  The quantities
`$A_1^+$(theo)' and `$A_2^+$(theo)' correspond to the amplitudes deduced from
$A_1^-$ and $A_2^-$ via Eq.~(\ref{eq:amplitudes}).  Apart from the $A_1^+$(theo)
for model \texttt{M6}, these values accurately reproduce the numerically
obtained amplitudes, $A_1^+$ and $A_2^+$, thus showing that the relationship on
amplitudes is respected.

\begin{table}[htbp]
\begin{center}
\caption{Wave vector components and amplitudes at the discontinuity for island modes
in three of the models.\label{tab:wave_parameters}}
\begin{tabular}{cccc}
\hline
\hline
& \texttt{M6} & \texttt{M7} & \texttt{M7b} \\
\hline
$\omega/\OmegaK$       & 15.679  & 15.843  &  16.276 \\
$c_+^2$                & 0.03083 & 0.01010 & 0.00201 \\
$c_-^2$                & 0.01584 & 0.00519 & 0.00391 \\
$\kpar$                & 30.480  &  7.457  &   5.177 \\
$\kperp^+$             & 83.935  & 157.434 & 362.831 \\
$\kperp^-$             & 120.782 & 219.736 & 260.074 \\
$\kpar/\kperp^+$       & -0.363  & -0.0474 & -0.0143 \\
$\kpar/\kperp^+$(rays) & -0.260  & -0.0369 & -0.0078 \\
$\kpar/\kperp^-$       & -0.252  & -0.0339 & -0.0199 \\
$\kpar/\kperp^-$(rays) & -0.184  & -0.0264 & -0.0109 \\
$A_1^-$                & -0.1475 &  0.1391 &  0.4057 \\
$A_2^-$                &  1.0000 &  1.0000 &  1.0000 \\
$A_1^+$                &  0.0061 &  0.2600 &  0.2888 \\
$A_1^+$(theo)          &  0.0019 &  0.2608 &  0.2884 \\
$A_2^+$                &  0.8463 &  0.8791 &  1.1169 \\
$A_2^+$(theo)          &  0.8505 &  0.8783 &  1.1173 \\
\hline
\end{tabular} \\
{\scriptsize The quantities $\kpar$ and $\kperp^{\pm}$ are the horizontal and
vertical components of the wave vectors.   The quantities $\kpar/\kperp^+$ and
$\kpar/\kperp^-$ are given a negative sign since the amplitudes $A_2^{\pm}$
(corresponding to the wave vectors $\vect{k}_2^{\pm} = \kpar
\vect{e}_{\parallel} - \kperp^{\pm} \vect{e}_{\perp}$) are larger (in absolute
value) than the amplitudes $A_1^{\pm}$.}
\end{center}
\end{table}

Another comparison carried out in Table~\ref{tab:wave_parameters} is between the
incidence/departure angles of the wave vectors and the predictions from ray
dynamics analysis.  The quantities $\kpar/\kperp^{\pm}$ correspond to the
numerically determined values of $\tan \vartheta^{\pm}$ where $\vartheta^{\pm}$
are the angles between the surface normal and the wave vector\footnote{Negative
values of $\vartheta$ simply mean that the wave is penetrating inwards (that is,
$r$ is decreasing) for increasing colatitudes, $\theta$.}.  The quantities
`$\kpar/\kperp^{\pm}$(rays)' are determined via ray dynamics.  A comparison
between the two shows some discrepancies but the values remain of the same
order. These differences are likely due to the limited accuracy of our approach
for extracting the wave vector components, and the fact that the mode behaviour
is more complex than what is predicted by ray dynamics.

Finally, as can be seen from the values of $A_1^{\pm}$, the amplitudes of the
secondary waves, although smaller, is not negligible.  Hence, these can be
expected to affect the phase shift that occurs at the discontinuity in the
primary wave and may possibly lead to increased coupling with other modes due to
the modified mode geometry, thereby leading to more avoided crossings.

\subsection{Frequency patterns}

\begin{figure*}[htbp]
\includegraphics[width=\columnwidth]{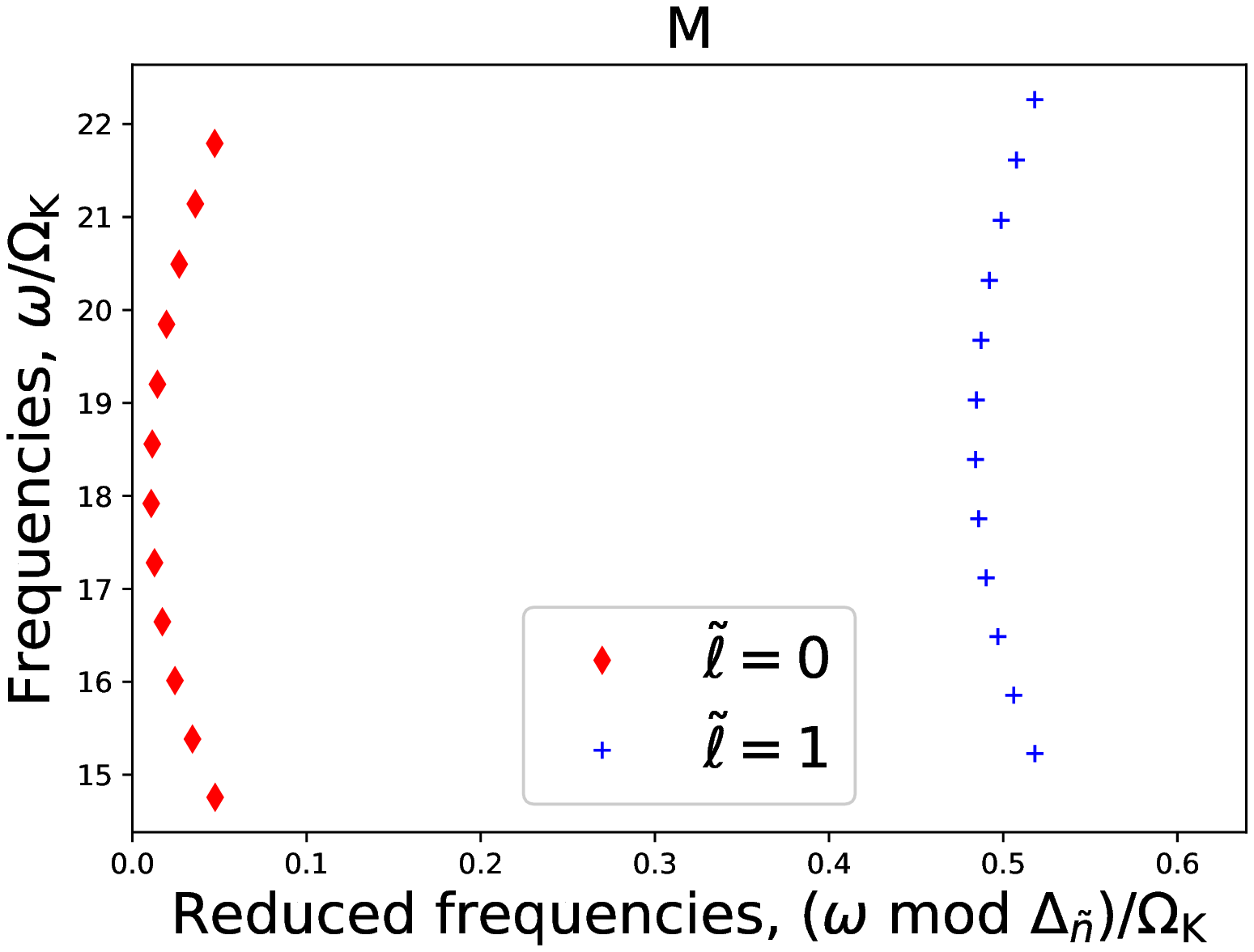} \hfill
\includegraphics[width=\columnwidth]{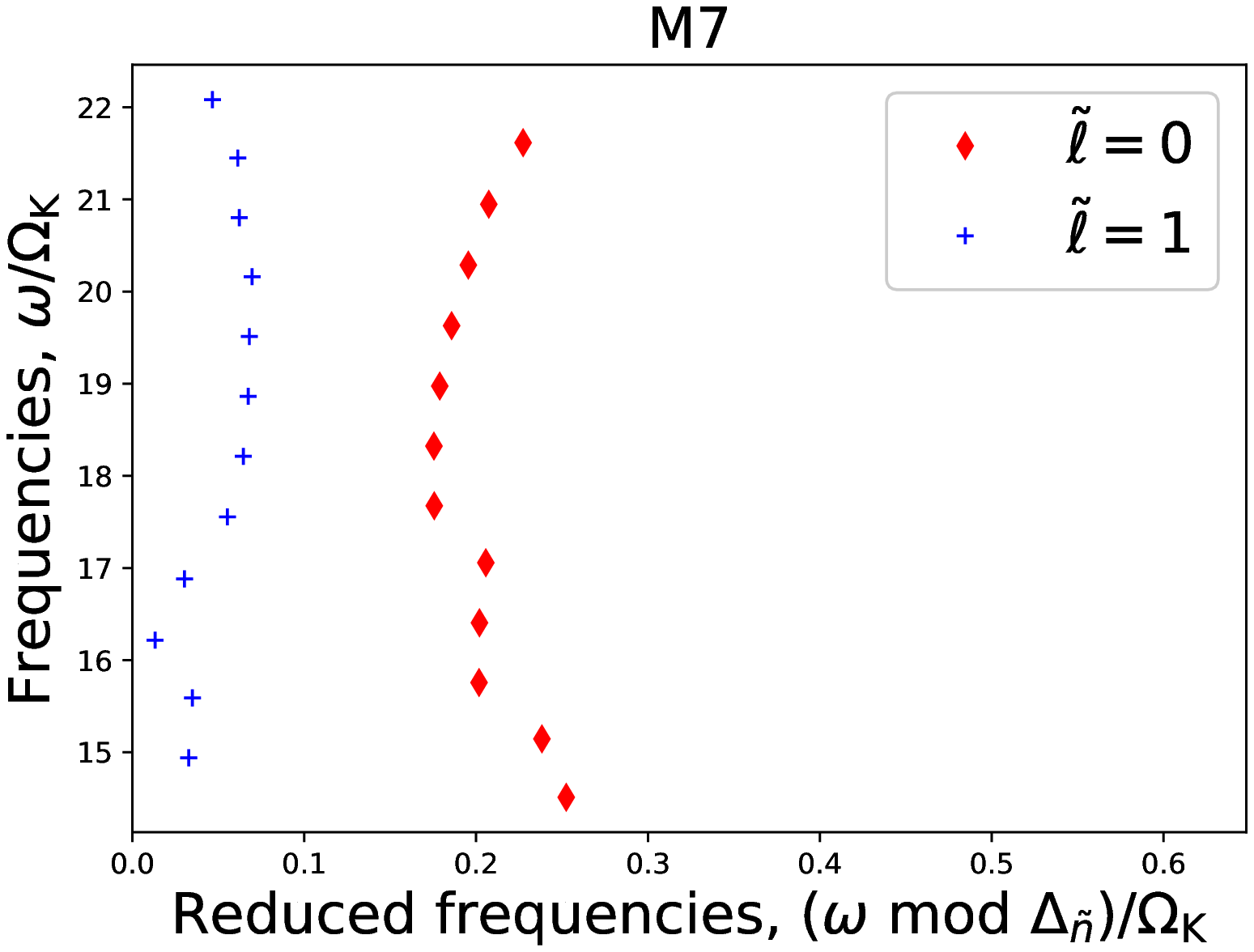} \\
\includegraphics[width=\columnwidth]{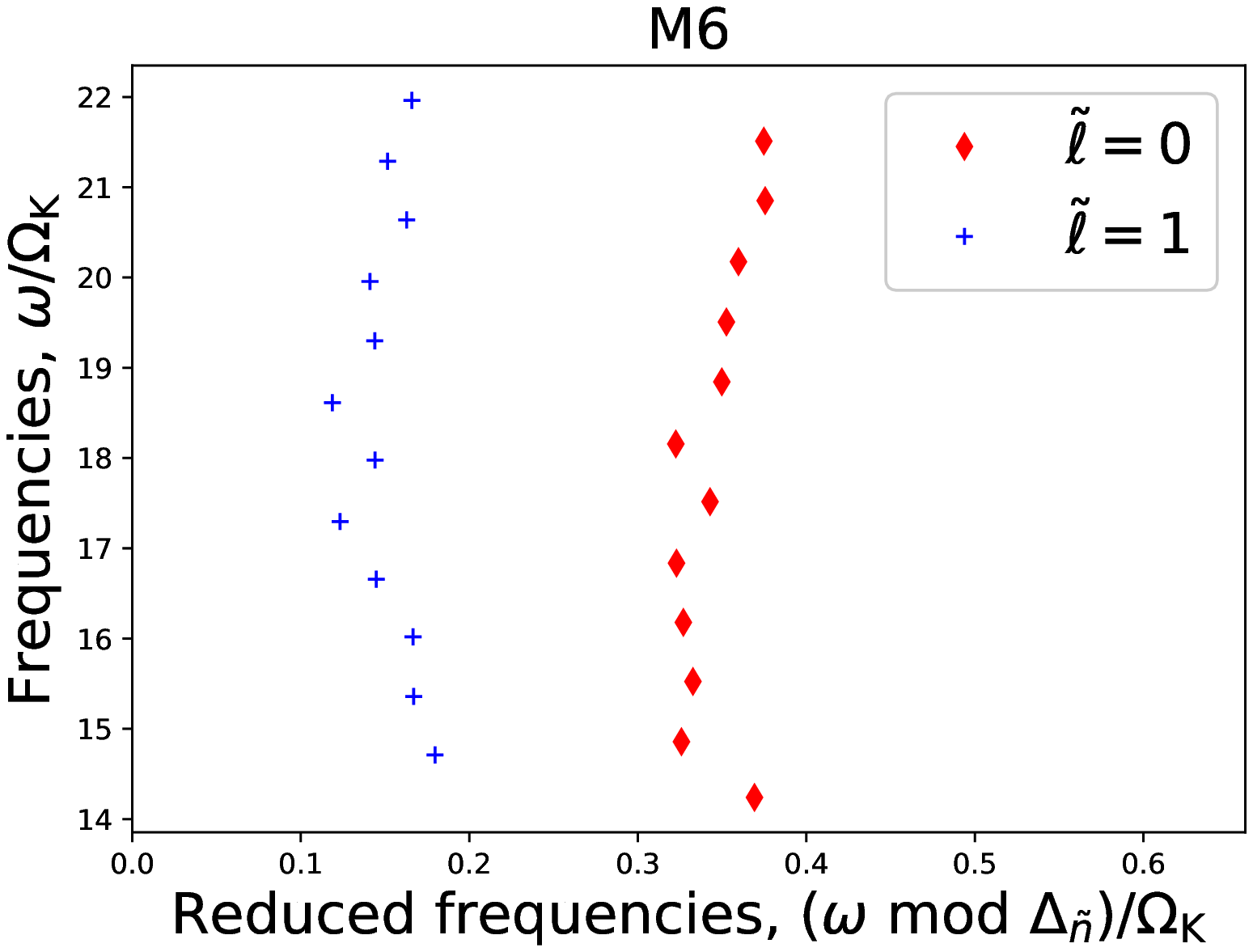} \hfill
\includegraphics[width=\columnwidth]{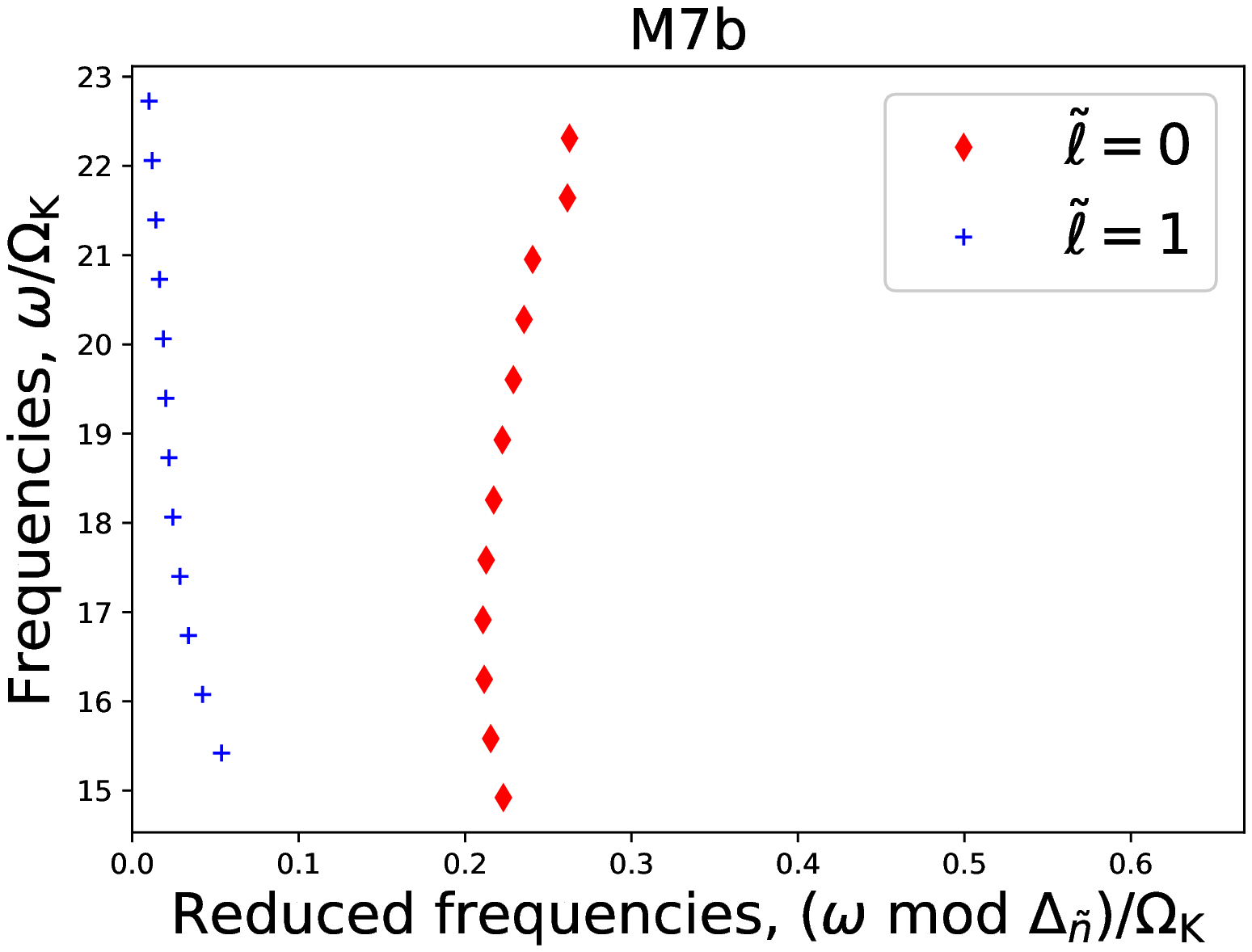}
\caption{Echelle diagrams for the axisymmetric modes in four of the models.
\label{fig:echelle}}
\end{figure*}

We now briefly address the question of whether discontinuities can adversely
affect frequency patterns to the point of hindering their detection in observed
stars.  A tool frequently used in solar-like stars is the so-called
echelle diagram \citep[\eg][]{Bedding2020}, in which the frequencies
are plotted as a function of the frequencies modulo the large separation.  Due
to the nearly equidistant frequency patterns in such stars, modes with the same
spherical degree line up on vertical ridges in echelle diagrams.  In
Fig.~\ref{fig:echelle}, we produce similar echelle diagrams using the pseudo
large separation, $\Delta_{\tilde{n}}$, and only plotting $m=0$ modes for the
sake of clarity.  Although the discontinuities lead to a more irregular
behaviour, clear ridges remain for the different $\tilde{\l}$ values. 
\citet{Reese2014} also reached a similar conclusion using histograms of
frequency differences for 3 M$_{\odot}$ models with discontinuities. 
Indeed, they found that the pseudo large separation could still be identified in
the discontinuous models.

We also carry out a more quantitative comparison between the pulsation
frequencies and a fit based on a simplified version of the asymptotic formula
for island mode frequencies \citep[\eg][]{Reese2009a}:
\begin{equation}
      \omega = \tilde{n} \Delta_{\tilde{n}}
             + \tilde{\l}\Delta_{\tilde{\l}}
             + m^2 \Delta_{\tilde{m}}
             - m \Omega_{\mathrm{fit}}
             + \tilde{\alpha},
\end{equation}
where $\Delta_{\tilde{n}}$, $\Delta_{\tilde{\l}}$, $\Delta_{\tilde{m}}$,
and $\tilde{\alpha}$ are various parameters related to the stellar structure
\citep{Lignieres2009,Pasek2012}, and $\Omega_{\mathrm{fit}}$  an average value
of the rotation rate appropriate for the set of modes under consideration.  We
therefore fit these parameters to reproduce the pulsation spectra of our models
for the same set of modes as described in Sect.~\ref{sect:variational}.
Table~\ref{tab:fit_accuracy} provides the root mean square differences and
maximal differences between the numerical and asymptotic frequencies, normalised
by the pseudo large separation, $\Delta_{\tilde{n}}$.  As expected, the
frequencies of model \texttt{M} are the closest to the asymptotic formula.  The
mean difference in the realistic model is intermediate between the best and
worst model.  In terms of maximal differences, model \texttt{Mreal} is among the
best whereas model \texttt{M6} is the worst model, very likely as a result of
the increased number of avoided crossings affecting the modes.  In all cases,
the differences are a few percent of the pseudo large separation (which itself
is half the classical large separation), meaning the frequency pattern is still
well-preserved and should be possible to identify with a suitable analysis.

\begin{table}[htbp]
\caption{Root mean square and maximal differences between numerical and
asymptotic frequencies for the different models.\label{tab:fit_accuracy}}
\begin{center}
\begin{tabular}{lccc}
\hline
\hline
\textbf{Model} & $\sqrt{\left<\delta\omega^2\right>}/\Delta_{\tilde{n}}$
               & $\max |\delta\omega|/\Delta_{\tilde{n}}$ \\
\hline
\texttt{M}     & 0.0198 & 0.0511 \\
\texttt{M6}    & 0.0282 & 0.1207 \\
\texttt{M7}    & 0.0326 & 0.0855 \\
\texttt{M7b}   & 0.0224 & 0.0663 \\
\texttt{Mreal} & 0.0273 & 0.0548 \\
\hline
\end{tabular}
\end{center}
\end{table}

Nonetheless, other factors may hinder finding the above frequency pattern. 
Indeed, the presence of chaotic modes with their own independent semi-random
frequency organisation \citep{Lignieres2009, Evano2019b}, or the lack of a clear
understanding of the mechanisms responsible for mode selection and pulsation
amplitudes both contribute to masking the frequency pattern associated with
acoustic island modes.

\section{Conclusion}

In this work, we calculated, thanks to an adiabatic version of the TOP code,
acoustic pulsation modes in rapidly rotating continuous and discontinuous
stellar models based on the ESTER code.  This allowed us to investigate various
topics namely the variational principle for general 2D rotation profiles in
discontinuous models, generalised rotational splittings, and acoustic glitches.
Some of the important results are:
\begin{enumerate}
\item Generalised rotational splittings are well approximated via weighted
      integrals of the rotation profile using rotation kernels deduced from the
      variational principle, except for specific cases where avoided crossings
      lead to discrepancies.  This raises the question as to how accurately the
      rotation profile can be recovered using inverse theory.  In a forthcoming
      article, we plan to investigate this question using a variety of different
      rotation profiles.  In this regard, the automatic mode classification
      algorithm described in \citet{Mirouh2019} can be used to efficiently
      identify pairs of prograde-retrograde modes.
\item Discontinuities alter the acoustic frequency patterns, but not to the
      point of preventing their detection in observed stars (especially taking
      into account the unrealistic nature of the discontinuities in our models),
      thus lending credence to recent detections of large frequency separations
      and ridges in echelle diagrams in $\delta$ Scuti stars
      \citep[\eg][]{GarciaHernandez2015, Bedding2020}. The modifications to the
      frequency spectrum leads to glitch patterns the periodicity of which can
      be calculated in a simple way.  Nonetheless, the presence of avoided
      crossings and possibly partial wave reflection at the discontinuity cause
      deviations from theoretical expectations in some cases. Accordingly, it
      may be possible to determine acoustic depths of sharp transitions using
      glitch patterns in observed frequencies.
\end{enumerate}

In a forthcoming work, we plan to investigate acoustic pulsations of ESTER
models using a non-adiabatic version of TOP.  This will allow us to investigate
other topics such as mode excitation and mode behaviour near the stellar
surface.

\begin{acknowledgements}
The authors thank François Lignières, Vincent Prat, and Pierre Houdayer for
useful and interesting discussions.  DRR acknowledges the support of the French
Agence Nationale de la Recherche (ANR) to the ESRR project under grant
ANR-16-CE31-0007 as well as financial support from the Programme National de
Physique Stellaire (PNPS) of the CNRS/INSU co-funded by the CEA and the CNES.
GMM acknowledges funding by the STFC consolidated grant ST/R000603/1. DRR, GMM,
MR, BP benefited from the hospitality of ISSI as part of the SoFAR team early on
in 2018 and 2019. This work was granted access to the HPC resources of IDRIS
under the allocation 2011-99992  made by GENCI (``Grand Equipement National de
Calcul Intensif'') and to HPC resources of Calmip under project P0107.
\end{acknowledgements}

\bibliographystyle{aa}
\bibliography{biblio}

\appendix

\section{Interface conditions}
\label{app:interface_conditions}

\subsection{Domain continuity}
\label{app:interface_displacement}

\begin{figure}[htbp]
\begin{center}
\includegraphics[width=0.7\columnwidth]{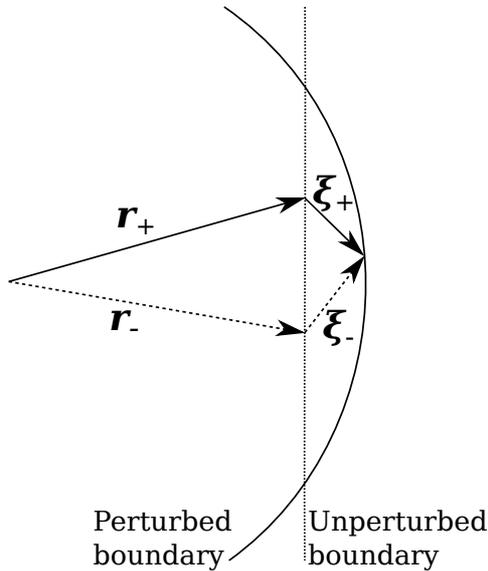}
\end{center}
\caption{Schematic illustration showing the movement of fluid parcels along
either side of a boundary that is perturbed by a pulsation mode.
\label{fig:perturbed_boundary}}
\end{figure}

In order to ensure that the domain remains continuous, one needs the boundary on
either side to be the same.  We will now consider a point on the perturbed
boundary. This point will be reached by fluid parcels on either side of the
boundary. The spatial coordinates of the fluid parcel just below the boundary
will be given by the formula:
\begin{equation}
\vect{r}_- + \vect{\xi}_-(\vect{r}_-,t).
\end{equation}
An analogous formula applies for the spatial coordinates just outside the
boundary. This leads to the following matching condition:
\begin{equation}
\label{eq:interface_displacement_appendix1}
\vect{r}_- + \vect{\xi}_-(\vect{r}_-,t) =
\vect{r}_+ + \vect{\xi}_+(\vect{r}_+,t).
\end{equation}
This matching relation is illustrated in Figure~\ref{fig:perturbed_boundary}. 
It is important to bear in mind that $\vect{r}_-$ is not necessarily equal to
$\vect{r}_+$, as illustrated in the figure, since the fluid may slip along
either side of the boundary. However, one can rearrange this expression as
follows:
\begin{equation}
\vect{r}_+ - \vect{r}_- =
\vect{\xi}_-(\vect{r}_-,t) -  \vect{\xi}_+(\vect{r}_+,t).
\end{equation}
Given that $\vect{\xi}_-$ and $\vect{\xi}_+$ are arbitrarily small, the
difference $\vect{r}_+ - \vect{r}_-$ is a vector tangent to the surface.  Hence,
calculating the dot product of the above equation and $\vect{n}$, the normal to
the surface, cancels out the left-hand side and yields:
\begin{equation}
\label{eq:interface_displacement_appendix2}
\vect{\xi}_-(\vect{r}_-,t)\cdot\vect{n} =   \vect{\xi}_+(\vect{r}_+,t)\cdot\vect{n}.
\end{equation}
The difference between $\vect{\xi}_+(\vect{r}_+,t)$ and
$\vect{\xi}_+(\vect{r}_-,t)$ is a second order term.  Hence, the above
condition reduces to Eq.~(\ref{eq:interface_displacement}).

\subsection{Condition on the pressure perturbation}
\label{app:interface_pressure}

The following condition ensures that the pressure remains continuous during the
oscillatory movements:
\begin{equation}
P_{\mathrm{Total}}^- (\vect{r}_- + \vect{\xi}_-,t) = P_{\mathrm{Total}}^+ (\vect{r}_+ + \vect{\xi}_+,t),
\end{equation}
where we have kept the same notation as above and where $P_{\mathrm{Total}}$ is
the total pressure (equilibrium $+$ perturbation).  This equation can be
developed as follows:
\begin{equation}
P_0^-(\vect{r}_-) + \delta P_-(\vect{r}_-,t) =
P_0^+(\vect{r}_+) + \delta P_+(\vect{r}_+,t).
\end{equation}
We can then use Eq.~(\ref{eq:interface_displacement_appendix1}) to develop, say,
the right-hand side:
\begin{eqnarray*}
P_0^+(\vect{r}_+) &+& \delta P_+(\vect{r}_+,t) \\
   &=& P_0^+(\vect{r}_- + \vect{\xi}_- - \vect{\xi}_+) + \delta P_+(\vect{r}_- + \vect{\xi}_- - \vect{\xi}_+,t) \\
   &\simeq& P_0^+(\vect{r}_-) + \left(\vect{\xi}_- - \vect{\xi}_+\right) \cdot \grad P_0^+ + \delta P_+(\vect{r}_-,t),
\end{eqnarray*}
where we have neglected second order terms on the third line.  Combining this
with the previous equation leads to:
\begin{equation}
\delta P_-(\vect{r}_-,t) - \delta P_+(\vect{r}_-,t) = \left(\vect{\xi}_- - \vect{\xi}_+\right) \cdot \grad P_0^+,
\end{equation}
where we have used the continuity of the equilibrium pressure.  If the boundary
coincides with an isobar, the right-hand side of the above equation cancels out
because the difference $\vect{\xi}_- - \vect{\xi}_+$ is within the boundary.  If
the boundary is not an isobar, then in normal circumstances (that is, when the
model is continuous), the difference $\vect{\xi}_- - \vect{\xi}_+$ will be
$\vect{0}$.  Either way, this leads to the final condition,
Eq.~(\ref{eq:interface_pressure}).  There are, however, cases where the 
right-hand side may not cancel, for instance at the boundary of a convective
core with a different chemical composition than the rest of the star.  In such a
situation, baroclinic flows are set up within the equilibrium model
\citep{EspinosaLara2013}, and probably require setting a specific condition
which takes these flows into account.  Nonetheless, it is interesting to note
that the above condition is in fact symmetric with respect to either side of the
boundary.  Indeed, the term $\left(\vect{\xi}_- - \vect{\xi}_+\right) \cdot\grad
P_0^+$ could be replaced by $\left(\vect{\xi}_- - \vect{\xi}_+\right) \cdot\grad
P_0^-$, since it only involves the gradient of the pressure along the boundary
and the pressure is continuous across the boundary.

\subsection{Condition on the perturbation to gravitational potential}
\label{app:interface_gravitational_potential}

In much the same way as the pressure, the gravitational potential and its
gradient are kept continuous through the following relations:
\begin{eqnarray}
\Psi_{\mathrm{Total}}^-(\vect{r}_- + \vect{\xi}_-,t) &=& \Psi_{\mathrm{Total}}^+(\vect{r}_+ + \vect{\xi}_+,t), \\
\grad \Psi_{\mathrm{Total}}^-(\vect{r}_- + \vect{\xi}_-,t) &=& \grad \Psi_{\mathrm{Total}}^+(\vect{r}_+ + \vect{\xi}_+,t),
\end{eqnarray}
where $\Psi_{\mathrm{Total}}$ is the total gravitational potential (equilibrium
$+$ perturbation).  At this point, however, we will take a different approach
than above since we are dealing with the Eulerian rather than Lagrangian
perturbation of the gravitational potential. Firstly, the sums $\vect{r}_+ +
\vect{\xi}_+$ can be replaced by $\vect{r}_- + \vect{\xi}_-$ or vice versa so as
to have the same arguments everywhere.  Therefore, in what follows we will use
the generic notation $\vect{r} + \vect{\xi}$ which can be arbitrarily chosen as
$\vect{r}_- + \vect{\xi}_-$ or $\vect{r}_+ + \vect{\xi}_+$.  Developing both
sides of both equations, making use of the continuity of the equilibrium
gravitational and its gradient to cancel zeroth order terms, and neglecting
second order terms lead to the following equations:
\begin{eqnarray}
\Psi_-(\vect{r},t) &+& \vect{\xi}(\vect{r},t) \cdot \grad \Psi_0^-(\vect{r},t)  \nonumber \\
\label{eq:interface_Psi_appendix1}
    &=& \Psi_+(\vect{r},t) + \vect{\xi}(\vect{r},t) \cdot \grad \Psi_0^+(\vect{r},t), \\
\grad \Psi_-(\vect{r},t) &+& \vect{\xi}(\vect{r},t) \cdot \grad \left( \grad \Psi_0^-(\vect{r},t) \right)  \nonumber \\
\label{eq:interface_Psi_appendix2}
&=& \grad \Psi_+(\vect{r},t) + \vect{\xi}(\vect{r},t) \cdot \grad \left( \grad \Psi_0^+(\vect{r},t) \right).
\end{eqnarray}
Given that $\grad \Psi_0$ is continuous, the first equation reduces to:
\begin{equation}
\label{eq:interface_Psi_appendix3}
\Psi_- = \Psi_+.
\end{equation}
In tensorial notation, the left-hand side of the second equation becomes:
\begin{eqnarray*}
\partial_i \Psi_- \vect{E}^i_- &+& \tilde{\xi}^j \partial_j \left( \vect{E}^k_- \partial_k \Psi_0^- \right) \\
&=& \partial_i \Psi_- \vect{E}^i_- + \tilde{\xi}^j \left[\partial_j \left( \vect{E}^k_- \right) \partial_k \Psi_0^- + \vect{E}^k_- \partial_{jk}^2 \Psi_0^- \right] \\
&=& \partial_i \Psi_- \vect{E}^i_- + \tilde{\xi}^j \partial_j \left( \vect{E}^k_- \right) \partial_k\Psi_0^- 
             +  \tilde{\xi}^j \partial_{ij}^2 \Psi_0^- \vect{E}^i_-,
\end{eqnarray*}
where $\vect{E}_i$ is the natural basis, $\tilde{\xi}$ the components of
$\vect{\xi}$ over that basis, and $\vect{E}^i$ the dual basis.  Calculating the
dot product of the above equation with $\vect{E}_i^-$ yields:
\begin{equation}
\partial_i \Psi_-  - \tilde{\xi}^j \Gammam_{ij}^k \partial_k\Psi_0^- 
             +  \tilde{\xi}^j \partial_{ij}^2 \Psi_0^-,
\end{equation}
where $\Gamma_{ij}^k = \partial_i \left( \vect{E}_j \right) \cdot \vect{E}^k =
-\partial_i \left( \vect{E}^k \right) \cdot \vect{E}_j$ is the Christoffel
symbol of the second kind.  With our choice of mapping, only
$\Gamma_{\zeta\zeta}^{\zeta}$ is discontinuous across the boundary, hence the
notation $\Gammam_{\zeta\zeta}^{\zeta}$ and $\Gammap_{\zeta\zeta}^{\zeta}$.  All
of the other geometric quantities ($\vect{E}_i$, $\vect{E}^i$, $\Gamma_{ij}^k$
with $(i,j,k) \neq (\zeta,\zeta,\zeta)$) are continuous.  Inserting this
expression into the left-hand side of Eq.~(\ref{eq:interface_Psi_appendix2}) and
a similar expression in the right-hand side, and simplifying out continuous terms
(geometric, $\Psi_0$, and $\grad \Psi_0$) yields the following three relations:
\begin{eqnarray}
\dz \Psi_-  &+& \left(\partial_{\zeta\zeta}^2 \Psi_0^- - \Gammam_{\zeta\zeta}^{\zeta} \dz \Psi_0\right) \tilde{\xi}^{\zeta}  \nonumber \\ &=&
\label{eq:interface_Psi_appendix4}
\dz \Psi_+   +  \left(\partial_{\zeta\zeta}^2 \Psi_0^+ - \Gammap_{\zeta\zeta}^{\zeta} \dz \Psi_0\right) \tilde{\xi}^{\zeta}, \\
\dt \Psi_- &=& \dt \Psi_+, \\
\dphi \Psi_- &=& \dphi \Psi_+,
\end{eqnarray}
where we have made use of the fact that $\partial_{ij}^2 \Psi_0$ is continuous
if $(i,j) \neq (\zeta,\zeta)$.  One will in fact notice that the latter two
equations are also a direct consequence of
Eq.~(\ref{eq:interface_Psi_appendix3}).

At this point, it is useful to introduce Poisson's equation in tensorial
notation:
\begin{eqnarray}
\Lambda \rho_0 &=& \lapl \Psi_0 = \div \grad \Psi_0 = \partial_i \left( \vect{E}^j \partial_j \Psi_0 \right) \cdot \vect{E}^i \nonumber \\
  &=& g^{ij} \partial_{ij}^2 \Psi_0 + \partial_i \left( \vect{E}^j \right) \cdot \vect{E}^i \partial_j \Psi_0 \nonumber \\
  &=& g^{ij} \partial_{ij}^2 \Psi_0 + \left[\partial_i \left( \vect{E}^j \right) \cdot \vect{E}_k\right] \left[\vect{E}^i \cdot \vect{E}^k\right] \partial_j \Psi_0 \nonumber \\
  &=& g^{ij} \partial_{ij}^2 \Psi_0 - \Gamma_{ik}^{j} g^{ik} \partial_j \Psi_0 \nonumber \\
  &=& g^{\zeta\zeta} \left( \partial^2_{\zeta\zeta} \Psi_0 - \Gamma_{\zeta\zeta}^{\zeta} \dz \Psi_0 \right) + R,
\end{eqnarray}
where $\Lambda = 4 \pi G$ or $4 \pi$ in the dimensional or dimensionless case,
respectively, $g^{ij} = \vect{E}^i \cdot \vect{E}^j$ is the contravariant
components of the metric tensor, and $R$ is a sum of terms which are
continuous across the boundary.  This last expression can then be used to
simplify Eq.~(\ref{eq:interface_Psi_appendix4}):
\begin{equation}
\partial_{\zeta} \Psi_- + \frac{\Lambda \rho_0^- - R_-}{g^{\zeta\zeta}} \tilde{\xi}^{\zeta} =
\partial_{\zeta} \Psi_+ + \frac{\Lambda \rho_0^+ - R_+}{g^{\zeta\zeta}} \tilde{\xi}^{\zeta}.
\end{equation}
The terms $R_-$ and $R_+$ cancel out since $R$ is continuous across the
boundary.  The remaining equation is then
\begin{equation}
\label{eq:interface_Psi_appendix5}
\partial_{\zeta} \Psi_- + \frac{\Lambda \rho_0^- \zeta^2 \rz}{r^2+\rt^2} \xi^{\zeta} =
\partial_{\zeta} \Psi_+ + \frac{\Lambda \rho_0^+ \zeta^2 \rz}{r^2+\rt^2} \xi^{\zeta},
\end{equation}
where we have introduced $\xi^{\zeta}$, the $\zeta$ component of $\vect{\xi}$ on
the alternate basis \citep[see, \eg\ Eq.~31 of][]{Reese2006}. At this point, it
is useful to recall that $\tilde{\xi}^{\zeta}$ and hence $\xi^{\zeta}$ are
continuous across the boundary (see Eq.~(\ref{eq:interface_displacement_appendix2})). 
Hence, using $\vect{r}_- + \vect{\xi}_-$ or $\vect{r}_+ + \vect{\xi}_+$ in
Eq.~(\ref{eq:interface_Psi_appendix1}) leads to the same results.

\section{Variational principle}
\label{app:variational_principle}

\subsection{General formula}

In order to derive the variational formula which relates pulsation frequencies
and their associated eigenfunctions, we start by calculating the dot product
between Euler's equation (Eq.~(\ref{eq:Euler})) and the product of the
equilibrium density and the complex conjugate of a second displacement field,
$\vect{\eta}^*$, which at this point can be different from $\vect{\xi}$, and
integrate the total over the stellar volume, $V$:
\begin{eqnarray}
0 &=& \mathop{\mathlarger{\int}}_{V} \mathop{\mathlarger{\mathlarger{\{}}} \vlp^2 \rho_0 \vect{\xi} \cdot \vect{\eta}^*
      \underbrace{- 2i\vlp \rho_0 \left(\vect{\Omega} \times \vect{\xi}\right)\cdot\vect{\eta}^*}_{I} \nonumber \\
  & & \underbrace{- \rho_0 \left[\vect{\Omega} \times \left( \vect{\Omega} \times \vect{\xi} \right)\right] \cdot \vect{\eta}^*}_{II}
      \underbrace{- \rho_0 \vect{\eta}^* \cdot \left[\vect{\xi} \cdot \grad \left( s \Omega^2 \vect{e}_s\right)\right]}_{III} \nonumber \\
  & & \underbrace{- P_0 \vect{\eta}^* \cdot \grad \left(\dP\right)}_{IV}
      \underbrace{+ \vect{\eta}^* \cdot \grad P_0 \left(\drho - \dP\right)}_{V} \nonumber \\
  & & \underbrace{- \rho_0 \vect{\eta}^* \cdot \grad \Psi}_{VI}
      \underbrace{+ \rho_0 \vect{\eta}^* \cdot \grad \left(\frac{\vect{\xi}\cdot\grad P_0}{\rho_0}\right)}_{VII} \nonumber \\
  & & \underbrace{+ \vect{\eta}^* \cdot \left[\frac{\left(\vect{\xi}\cdot\grad P_0\right)\grad\rho_0 -
      \left(\vect{\xi}\cdot\grad\rho_0\right)\grad P_0}{\rho_0}\right]}_{VIII} \mathop{\mathlarger{\mathlarger{\}}}} \mathrm{dV}.
\end{eqnarray}
At this point, the goal is to reformulate the above integral so that it is
manifestly symmetric (in a Hermitian sense) with respect to $\vect{\xi}$ and
$\vect{\eta}$.  In what follows, it is very important to bear in mind that
$\Omega$ and its associated vector $\vect{\Omega} = \Omega \vect{e}_z$ depend on
$\zeta$ and $\theta$.  This is different than the approach taken in
\citet{Lynden-Bell1967} where $\Omega$ is constant (differential rotation is,
instead, taken into account as a background velocity field, $\vect{v}_0$). 
Furthermore, given that the equilibrium model may be discontinuous, it will be
important, for some of the terms, to decompose the stellar volume into
subdomains, $V_i$, such that the model is continuous in each subdomain. 
Obviously, the relation $V = \cup_i V_i$ holds. Finally, when dealing with the
gravitational potential, we will introduce the notation $V_e$ to represent an
external domain which comprises all of the space outside the star, and
$V_{\infty}$ to represent all of space, including the star.

Terms $I$ and $II$ can easily be rearranged into the following symmetric forms:
\begin{eqnarray}
I  &=& \int_V - 2i\vlp \rho_0 \vect{\Omega} \cdot \left( \vect{\xi} \times \vect{\eta}^* \right) \mathrm{dV}, \\
II &=& \int_V \left\{ -\rho_0 \left( \vect{\Omega} \cdot \vect{\xi} \right) \left( \vect{\Omega} \cdot \vect{\eta}^* \right) 
      + \rho_0 \Omega^2 \vect{\xi} \cdot \vect{\eta}^* \right\} \mathrm{dV}.
\end{eqnarray}

Term $III$ can be rewritten as:
\begin{eqnarray}
III &=& -\int_V \rho_0 \vect{\eta}^* \cdot \left[ \vect{\xi} \cdot \grad \left( \frac{\grad P_0}{\rho_0} + \grad \Psi_0\right) \right] \mathrm{dV} \nonumber \\
    &=& \sum_i \int_{V_i} \left\{ \frac{\left(\vect{\xi} \cdot \grad \rho_0\right) \left(\vect{\eta}^* \cdot \grad P_0\right)}{\rho_0}
     - \vect{\eta}^* \cdot \left[ \vect{\xi} \cdot \grad \left( \grad P_0 \right) \right] \right. \nonumber \\
    & & \left.  - \rho_0 \vect{\eta}^* \cdot \left[ \vect{\xi} \cdot \grad \left( \grad \Psi_0 \right) \right] \right\} \mathrm{dV},
\end{eqnarray}
where we have made use of the hydrostatic equilibrium equation, in which we have
neglected viscosity and meridional circulation.  It is important to note that on
the second and third line, the integral is carried out over $\sum_i V_i$. The
reason for this is that $\grad P_0$ may be discontinuous, meaning that $\grad
\left( \grad P_0 \right)$ has to be calculated over each separate subdomain,
$V_i$. The last two terms are symmetric.  This can be seen, for instance, by
expressing them explicitly in terms of their Cartesian coordinates: 
$\vect{\eta}^* \cdot \left[ \vect{\xi} \cdot \grad \left( \grad P_0 \right)
\right] = \left(\eta^i\right)^* \xi^j \partial_{ij}^2 P_0$.  The first term
cancels out with the last part of term $VIII$.

When developing term $IV$, it is important to treat each domain, $V_i$,
separately:
\begin{eqnarray}
IV &=& \sum_i \int_{V_i} \left\{ - \grad \cdot \left( P_0 \vect{\eta}^* \dP \right)
    + \dP \grad \cdot \left( P_0 \vect{\eta}^* \right) \right\} \mathrm{dV} \nonumber \\
   &=& -\sum_i \int_{B_i} \delta P \vect{\eta}^* \cdot \vect{\mathrm{dS}}
    + \sum_i \int_{V_i} \dP \div \left( P_0 \vect{\eta}^*\right) \mathrm{dV},
\end{eqnarray}
where $B_i$ denotes the bounds of subdomain $V_i$.  It turns out that the
surface terms cancel out.  Indeed, there are two possible cases.  In the first
case, the surface corresponds to an internal discontinuity. As explained in the
previous section, both the normal component of the displacement and the
Lagrangian pressure perturbation remain continuous across the discontinuity. 
Furthermore, there will be two surface terms, one for the domain just below the
discontinuity and  the other for the domain just above.  The vector
$\vect{\mathrm{dS}}$ takes opposite signs in both surface terms since it is
directed outwards from the domain.  As a result, the two terms cancel.  In the
second case, the surface term corresponds to the stellar surface.  As explained
in Sect.~\ref{sect:boundary_conditions}, we impose the simple mechanical
boundary condition $\delta p = 0$, thereby cancelling this surface term.

Terms $IV$ and $V$ may be combined as follows:
\begin{eqnarray}
IV + V &=&  \sum_i \int_{V_i} \left[\left(\vect{\eta}^* \cdot \grad P_0 \right) \left(\drho - \dP \right) \right. \nonumber \\
       & & \left. + \left(\vect{\eta}^* \cdot \grad P_0 + P_0 \div \vect{\eta}^*\right) \dP \right] \mathrm{dV} \nonumber \\
       &=&  \sum_i \int_{V_i} \left[\left(\vect{\eta}^* \cdot \grad P_0 \right) \drho 
          - \frac{\delta\tilde{\rho}^* \delta P}{\rho_0} \right] \mathrm{dV},
\end{eqnarray}
where we have made use of the continuity equation and introduced the Lagrangian
density perturbation, $\delta \tilde{\rho}$ associated with $\vect{\eta}$.

Term $VII$ is treated as follows:
\begin{eqnarray}
VII &=& \sum_i \int_{V_i} \left[ \div \left( \rho_0 \vect{\eta}^* \frac{\vect{\xi} \cdot \grad P_0}{\rho_0} \right)
        - \frac{\vect{\xi} \cdot \grad P_0}{\rho_0} \div \left(\rho_0 \vect{\eta}^* \right) \right] \mathrm{dV} \nonumber \\
    &=& \sum_i \int_{B_i} \left(\vect{\xi} \cdot \grad P_0\right) \vect{\eta}^* \cdot \vect{\mathrm{dS}}
      + \sum_i \int_{V_i} \left[\left(\vect{\xi} \cdot \grad P_0\right) \frac{\delta\tilde{\rho}^*}{\rho_0} \right. \nonumber \\
    & & \left. - \frac{\left(\vect{\xi} \cdot \grad P_0 \right)\left(\vect{\eta}^* \cdot \grad \rho_0\right)}{\rho_0} \right] \mathrm{dV},
\end{eqnarray}
where we have once more made use of the continuity equation. In the above
calculations, the surface terms do not cancel, but they are symmetric since both
the discontinuities and the stellar surface follow isobars. The term on the last
line cancels out with the first part of term $VIII$.

Last but not least, we deal with term $VI$.  As has been shown in
\citet{Unno1989} and \citet{Reese2006PhD}, this term can be rearranged into an
integral of the form $\frac{1}{\Lambda} \int_{V_{\infty}} \grad \Psi \cdot \grad
\Phi^* \mathrm{dV}$, where $\Lambda = 4\pi G$ or $4\pi$ in the dimensionless
case, and $\Phi$ is the gravitational potential associated with the displacement
field $\vect{\eta}$. However, surface terms and terms arising from internal
discontinuities were not dealt with in the above works.  In what follows, we
re-derive this  expression, while keeping track of such terms:
\begin{eqnarray}
VI &=& - \sum_i \int_{V_i} \div \left( \rho_0 \vect{\eta}^* \Psi \right) \mathrm{dV}
       + \sum_i \int_{V_i} \Psi \div \left( \rho_0 \vect{\eta}^* \right) \mathrm{dV} \nonumber \\
   &=& - \sum_i \int_{B_i}  \rho_0  \Psi \vect{\eta}^* \cdot \vect{\mathrm{dS}}
       - \sum_i \int_{V_i} \Psi \tilde{\rho}^* \mathrm{dV} \nonumber \\
   &=& - \sum_{i+e} \int_{B_i}  \rho_0  \Psi \vect{\eta}^* \cdot \vect{\mathrm{dS}}
       - \frac{1}{\Lambda}\sum_{i+e} \int_{V_i} \Psi \lapl \Phi^* \mathrm{dV} \nonumber \\
   &=& - \sum_{i+e} \int_{B_i}  \rho_0  \Psi \vect{\eta}^* \cdot \vect{\mathrm{dS}}
       - \frac{1}{\Lambda}\sum_{i+e} \int_{V_i} \div \left( \Psi \grad \Phi^* \right) \mathrm{dV} \nonumber \\
   & & + \frac{1}{\Lambda}\sum_{i+e} \int_{V_i} \grad \Psi \cdot \grad \Phi^* \mathrm{dV} \nonumber \\
   &=& - \sum_{i+e} \int_{B_i} \underbrace{\Psi \left(\rho_0 \vect{\eta}^*  + \frac{\grad \Phi^*}{\Lambda} \right) \cdot \vect{\mathrm{dS}}}_{(a)}
       + \frac{1}{\Lambda}\sum_{i+e} \int_{V_i} \grad \Psi \cdot \grad \Phi^* \mathrm{dV} \nonumber \\
   &=& \frac{1}{\Lambda}\sum_{i+e} \int_{V_i} \grad \Psi \cdot \grad \Phi^* \mathrm{dV},
\end{eqnarray}
where $\tilde{\rho}$ is the Eulerian density perturbation associated with
$\vect{\eta}$, the notation `$i+e$' stands for internal domains plus the
external domain $V_e$. Various steps in the above developments need further
explanation.  Firstly, on the third line, the external domain was incorporated
along with internal domains.  This step is justified because the supplementary
terms are equal to zero.  It must be noted that the surface associated with
$V_e$, only includes the lower bound, that is, the stellar surface.  Secondly,
the divergence (or Ostrogradsky's) theorem was used to transform volume
integrals on lines one and four into surface integrals. While straightforward in
most cases, it is not as obvious on line four for the external domain $V_e$. 
Indeed, it is not clear if some external boundary at infinity should be included
or not. This problem can be dealt with in a rigorous way by considering an
external domain, $\tilde{V}_e$, which is bounded by a sphere of radius $R_e$ and
then taking the limit as $R_e$ goes to infinity.  Such an approach was taken in
\citet{Reese2006PhD} who showed that the external surface term goes to zero in
such conditions.  Finally, it is necessary to show that the surface terms cancel
out.  We first start by noting that the surface element, $\vect{\mathrm{dS}}$,
is parallel to the vector $\vect{E}^{\zeta}$, and so may be written as
$\tilde{\mathrm{dS}} \vect{E}^{\zeta}$.  Hence, the integrand in the surface
terms may be written:
\begin{eqnarray}
(a) &=& -\Psi \left[\rho_0 \vect{\eta}^* + \frac{\grad \Phi^*}{\Lambda} \right] \cdot \vect{E}^{\zeta} \tilde{\mathrm{dS}} \nonumber \\
    &=& -\Psi \left[\rho_0 \left(\eta^{\zeta}\right)^* + \frac{g^{\zeta j}\partial_j \Phi^*}{\Lambda}\right] \tilde{\mathrm{dS}} \nonumber \\
    &=& -\Psi \left[\frac{\gzz}{\Lambda} \left(\dz \Phi^* + \frac{\Lambda\rho_0\left(\eta^{\zeta}\right)^*}{\gzz}\right)
        + \frac{g^{\zeta,j\neq\zeta}\partial_{j\neq\zeta}\Phi^*}{\Lambda}\right] \tilde{\mathrm{dS}}.
\end{eqnarray}
Now, we recall that each internal boundary (either from a discontinuity or from
the stellar surface) gives rise to two surface terms, one from the domain just
below the boundary and the other from the domain just above.  As can be seen in
the above expression, these two surface terms will cancel out.  Indeed, based on
the interface conditions (Eqs.~(\ref{eq:interface_Psi})
and~(\ref{eq:interface_grad_Psi})) and the continuity of $g^{ij}$ (for our
choice of coordinate system), the first part of the above expression is
continuous across the boundary.  Only $\tilde{\mathrm{dS}}$ changes signs given
that the vector $\vect{\mathrm{dS}}$ is always directed outwards from the
corresponding domain.

We now combine the above formulas to obtain the following expression:
\begin{eqnarray}
0 &=& \sum_i \mathop{\mathlarger{\int}}_{V_i} \mathop{\mathlarger{\mathlarger{\mathlarger{\mathlarger{\{}}}}} \vlp^2 \rho_0 \vect{\xi} \cdot \vect{\eta}^*
      - 2i\vlp \rho_0 \vect{\Omega} \cdot \left( \vect{\xi} \times \vect{\eta}^* \right) \nonumber \\
  & & -\rho_0 \left( \vect{\Omega} \cdot \vect{\xi} \right) \left( \vect{\Omega} \cdot \vect{\eta}^* \right) 
      + \rho_0 \Omega^2 \vect{\xi} \cdot \vect{\eta}^*
      - \vect{\eta}^* \cdot \left[ \vect{\xi} \cdot \grad \left( \grad P_0 \right) \right] \nonumber\\
  & & - \rho_0 \vect{\eta}^* \cdot \left[ \vect{\xi} \cdot \grad \left( \grad \Psi_0 \right) \right]
      + \left(\vect{\eta}^*\cdot\grad P_0\right)\drho
      + \left(\vect{\xi}\cdot\grad P_0\right)\frac{\delta\tilde{\rho}^*}{\rho_0} \nonumber \\
  & & - \frac{\delta p \delta\tilde{\rho}^*}{\rho_0} \mathop{\mathlarger{\mathlarger{\mathlarger{\mathlarger{\}}}}}} \mathrm{dV}
      + \sum_i \int_{B_i} \left(\vect{\xi} \cdot \grad P_0\right) \vect{\eta}^* \cdot \vect{\mathrm{dS}} \nonumber \\
  & & - \int_S \delta P \vect{\eta}^* \cdot \vect{\mathrm{dS}}
      + \frac{1}{\Lambda}\sum_{i+e} \int_{V_i} \grad \Psi \cdot \grad \Phi^* \mathrm{dV}.
\end{eqnarray}
We note that at this stage, we have not yet made use of the adiabatic
approximation (apart from cancelling out a surface term, thanks to the boundary
condition $\delta p = 0$, a condition which is usually applied in adiabatic
calculations).  We now use the adiabatic relation, $\drho = \frac{1}{\Gamma_1}
\dP$, to replace the Lagrangian density variations by Lagrangian pressure
perturbations, and then develop these in terms of Eulerian pressure perturbations
and displacement fields.  The final result is:
\begin{eqnarray}
0 &=& \sum_i \mathop{\mathlarger{\int}}_{V_i} \mathop{\mathlarger{\mathlarger{\mathlarger{\mathlarger{\{}}}}} \vlp^2 \rho_0 \vect{\xi} \cdot \vect{\eta}^*
      - 2i\vlp \rho_0 \vect{\Omega} \cdot \left( \vect{\xi} \times \vect{\eta}^* \right) \nonumber \\
  & & -\rho_0 \left( \vect{\Omega} \cdot \vect{\xi} \right) \left( \vect{\Omega} \cdot \vect{\eta}^* \right) 
      + \rho_0 \Omega^2 \vect{\xi} \cdot \vect{\eta}^*
      - \vect{\eta}^* \cdot \left[ \vect{\xi} \cdot \grad \left( \grad P_0 \right) \right] \nonumber\\
  & & - \rho_0 \vect{\eta}^* \cdot \left[ \vect{\xi} \cdot \grad \left( \grad \Psi_0 \right) \right]
      - \frac{\pi^* P}{\Gamma_1 P_0}
      + \frac{\left(\vect{\xi} \cdot \grad P_0\right)\left(\vect{\eta}^* \cdot \grad P_0\right)}{\Gamma_1 P_0} \mathop{\mathlarger{\mathlarger{\mathlarger{\mathlarger{\}}}}}} \mathrm{dV} \nonumber \\
  & & + \sum_i \int_{B_i} \left(\vect{\xi} \cdot \grad P_0\right) \vect{\eta}^* \cdot \vect{\mathrm{dS}}
      + \frac{1}{\Lambda}\sum_{i+e} \int_{V_i} \grad \Psi \cdot \grad \Phi^* \mathrm{dV},
\label{eq:variational_appendix}
\end{eqnarray}
where $\pi$ is the Eulerian pressure perturbation associated with the
displacement fields $\vect{\eta}$. This expression is manifestly symmetric in
$(\vect{\xi},P,\Psi)$ and $(\vect{\eta},\pi,\Phi)$ and consequently leads to the
variational principle \citep{Lynden-Bell1967}.  A useful consequence of this is
the quadratic convergence of the variational frequencies (obtained by assuming
that $(\vect{\xi},P,\Psi)=(\vect{\eta},\pi,\Phi)$ and solving the above equation
for $\omega$) to the true frequency, as a function of the error on the
eigenfunctions \citep[\eg][]{Christensen-Dalsgaard1982}.

\subsection{Explicit formulas}

In what follows, we provide explicit expressions for the different terms
which intervene in Eq.~(\ref{eq:variational_appendix}), a number of which
were already given in \citet{Reese2006}.  Such expressions are needed when
evaluating numerically the variational frequency:
\begin{eqnarray}
\| \vect{\xi} \|^2 &=&  |\xiz|^2 \frac{\zeta^4}{r^4} + |\xit|^2 \frac{\zeta^2(r^2+\rt^2)}{r^4 \rz^2}
                       +|\xip|^2\frac{\zeta^2}{r^2\rz^2} \nonumber \\
                   & &+ 2 \Re \left\{ \left(\xiz\right)^*\xit\right\} \frac{\zeta^3\rt}{r^4\rz}, \\
i\vect{\Omega} \cdot \left( \vect{\xi} \times \vect{\xi}^* \right)
                   &=& 2 \Omega\left[ \left( \frac{\cost}{\rz} + \frac{\rt\sint}{r\rz}
                         \right) \frac{\zeta^2\left(\xi_r^{\theta} \xi_i^{\phi} - \xi_r^{\phi}
                         \xi_i^{\theta} \right)}{r^2\rz} \right. \nonumber \\
                   & & \left. + \frac{\zeta^3 \sint \left(
                       \xi_r^{\zeta} \xi_i^{\phi}-\xi_r^{\phi}\xi_i^{\zeta} \right)}{r^3\rz} \right], \\
\| \grad \Psi \|^2 &=& \frac{r^2 + \rt^2}{r^2\rz^2} \left|
                       \dz \Psi \right|^2  + \frac{1}{r^2} \left| \dt \Psi \right|^2
                      +\frac{1}{r^2 \sin^2\theta} \left| \dphi \Psi \right|^2 \nonumber \\
                   & &- \frac{2\rt}{r^2\rz} \Re \left( \dz \Psi^* \dt \Psi \right), \\
\left| \vect{\Omega} \cdot \vect{\xi} \right|  &=& \Omega^2
       \left|\frac{\zeta^2\rz\cost}{r^2\rz}\xiz + \frac{\zeta\left(\rt\cost-r\sint\right)}{r^2\rz}\xit\right|^2.
\end{eqnarray}

The term $\vect{\xi}^* \cdot \left[\vect{\xi} \cdot \grad \left(\grad P_0
\right) \right]$ can be developed through tensor analysis:
\begin{eqnarray}
\vect{\xi}^* &\cdot& \left[\vect{\xi} \cdot \grad \left(\grad P_0 \right) \right] \nonumber \\
  &=& \left(\tilde{\xi}^j\right)^* \vect{E}_j \cdot \left\{\tilde{\xi}^i \d_i \left[\left(\d_k P_0\right) \vect{E}^k\right]\right\} \nonumber \\
  &=& \left(\tilde{\xi}^j\right)^* \vect{E}_j \cdot \left[\tilde{\xi}^i \left(\d_{ik}^2 P_0\right) \vect{E}^k + \tilde{\xi}^i \left(\d_k P_0\right) \left(\d_i \vect{E}^k\right)\right] \nonumber \\
  &=& \left(\tilde{\xi}^j\right)^* \tilde{\xi}^i \left(\d_{ij}^2 P_0 - \Gamma_{ij}^k\partial_k P_0\right) \nonumber \\
  &=& \frac{\zeta^4}{r^4\rz^2} \left(\dzz P_0 - \frac{\rzz}{\rz}\dz P_0\right) \left|\xiz\right|^2 \nonumber \\
  & & +\frac{2\zeta^3}{r^4\rz^2} \left[\dzt P_0 - \left( \frac{\rzt}{\rz} - \frac{\rt}{r} \right)\dz P_0 - \frac{\rz}{r}\dt P_0\right] \Re\left[\left(\xiz\right)^*\xit\right] \nonumber \\
  & & +\frac{\zeta^2}{r^4\rz^2}\left(\dtt P_0 - \frac{r\rtt-2\rt^2-r^2}{r\rz}\dz P_0 - \frac{2\rt}{r}\dt P_0\right) \left|\xit\right|^2 \nonumber \\
  & & +\frac{\zeta^2}{r^4\rz^2}\left(\frac{r-\rt\cott}{\rz}\dz P_0 + \cott\dt P_0\right)\left|\xip\right|^2,
\end{eqnarray}
where we have expressed the displacement using the alternate components on the
last line.

\section{Ray dynamics}
\label{app:ray_dynamics}

Ray trajectories were calculated in the usual spherical coordinate system
$(r,\theta,\phi)$ using the system of equations provided in \citet{Prat2016}.  
We used $H = \omega^2 = c_0^2 k^2$ as the Hamiltonian function. This leads to
the following system:
\begin{eqnarray}
\label{eq:rays1}
\dfrac{r}{t} &=& \dpart{H}{k_r} = 2 c_0^2 k_r, \\
\label{eq:rays2}
\dfrac{\theta}{t} &=& \frac{1}{r} \dpart{H}{\kt} = \frac{2 c_0^2 \kt}{r}, \\
\label{eq:rays3}
\dfrac{k_r}{t}    &=& -\dpart{H}{r} + \frac{\kt}{r} \dpart{H}{\kt}
                   =  -k^2 \left(\dpart{c_0^2}{r}\right)_{\theta} + \frac{2 c_0^2 \kt^2}{r}, \\
\label{eq:rays4}
\dfrac{\kt}{t}    &=& -\frac{1}{r}\dpart{H}{\theta} + \frac{\kt}{r}\dpart{H}{k_r}
                   =  -\frac{k^2}{r}\left(\dpart{c_0^2}{\theta}\right)_r + \frac{2 c_0^2 k_r \kt}{r}.
\end{eqnarray}
Although the ray trajectories are calculated in the spherical coordinate system,
the various derivatives of $c_0^2$ are first calculated in the spheroidal
coordinate system before being converted to the spherical system via the
following relations:
\begin{eqnarray}
\left(\dpart{c_0^2}{r}\right)_{\theta} &=& \frac{1}{\rz}\left(\dpart{c_0^2}{\zeta}\right)_{\theta}, \\
\left(\dpart{c_0^2}{\theta}\right)_{r} &=& \left(\dpart{c_0^2}{\theta}\right)_{\zeta}
                                        -  \frac{\rt}{\rz}\left(\dpart{c_0^2}{\zeta}\right)_{\theta}.
\end{eqnarray}
The system of equations~(\ref{eq:rays1}) to~(\ref{eq:rays4}) is solved
numerically for an initial position and wave vector using a fourth order
Runge-Kutta method, except near discontinuities and the stellar surface where
Heun's third-order method (also a Runge-Kutta method) is used instead. Indeed at
these locations, the step is adjusted so as to fall precisely on the relevant
boundary, thus making it easier to apply a wave reflection or Snell-Descartes'
law, and the use of Heun's method reduces the risk of overstepping this
boundary, unlike the fourth order Runge-Kutta method.

\section{Toy model for glitches}
\label{app:toy_model}

\begin{figure}[htbp]
\includegraphics[width=\columnwidth]{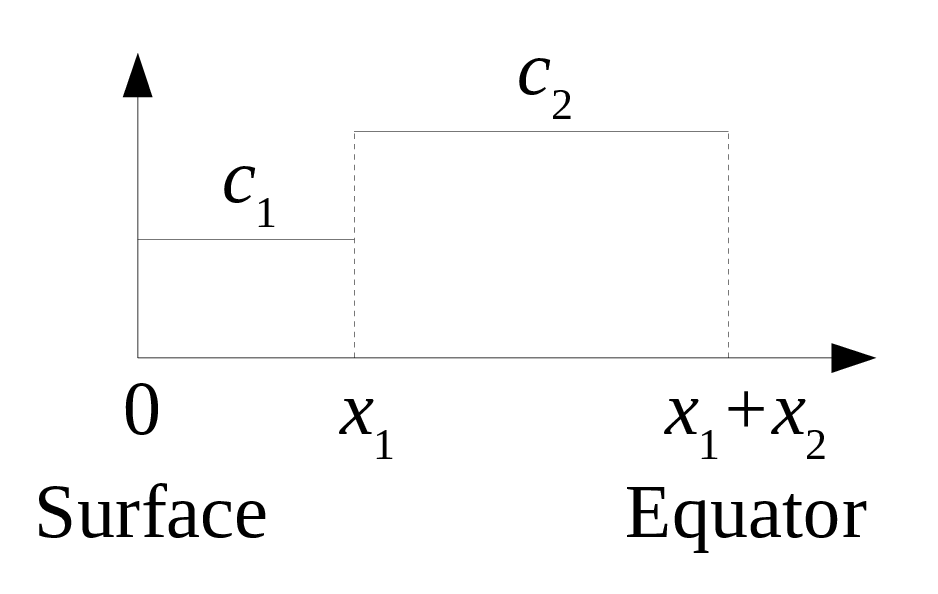}
\caption{Sound velocity profile in toy model.  Only half the model is shown, the
other half (beyond the centre, that is, $x_1 + x_2$) being symmetric.
\label{fig:toy_model}}
\end{figure}

In this section, we consider a 1D toy model representative of a sound wave
travelling along an island mode ray path in the presence of a discontinuity. 
Figure~\ref{fig:toy_model} illustrates half of this model, the other half being
deduced by symmetry.  For the sake of simplicity, constant sound velocities,
denoted $c_1$ and $c_2$, are used over the domains $[0,x_1[$ and $[x_1,\xT]$ (as
well as their symmetric counterparts), where $\xT = x_1 + x_2$.  We assume that
the density is discontinuous between the two domains whereas the pressure
and first adiabatic exponent are continuous, in accordance with our stellar
models.

We then consider the following set of simplified pulsation equations:
\begin{eqnarray}
0 &=& \drho + \dfrac{\xi}{x}, \\
-\omega^2 \xi_x &=& - \frac{P_0}{\rho_0} \dfrac{}{x} \dP, \\
0 &=& \dP - \Gamma_1 \drho,
\end{eqnarray}
along with the boundary conditions:
\begin{equation}
\dP(x=0) = 0
\end{equation}
at the surface and
\begin{equation}
\dP(x=x_T) = 0 \qquad \mbox{or} \qquad \dfrac{}{x} \dP (x=x_T) = 0
\end{equation}
at the equator (that is, at $\xT$).  The latter conditions come from the fact
that pulsation modes are either antisymmetric or symmetric with respect to the
equator.  Finally, the following interface conditions apply at $x=x_1$:
\begin{equation}
\dP(x=x_1^{-}) = \dP(x=x_1^{+}), \qquad \xi_x(x=x_1^-) = \xi_x(x=x_1^+).
\end{equation}

The pressure perturbation then takes on the following form:
\begin{equation}
\dP = \left\{
\begin{array}{ll}
A_1 \sin\left(k_1 x\right) &\mbox{for} \quad x \in [0, x_1[, \\
A_2 \sin\left(k_2 (x-\xT)\right) \quad\mbox{or} \quad & \multirow{2}{*}{for $\quad x \in [x_1,\xT]$,} \\
\quad A_2 \cos\left(k_2(x-\xT)\right)\quad &
\end{array}
\right.
\end{equation}
where $k_i = \omega/c_i$.  The two options for the solution in the $[x_1, \xT]$
domain correspond to antisymmetric (or odd) and symmetric (or even) solutions,
respectively.  Enforcing the interface condition then leads to the following
discriminants which define the eigenvalues:
\begin{equation}
\label{eq:toy_model_discriminant1}
\frac{\sin(\omega \tau_1)\cos(\omega \tau_2)}{k_2} + 
\frac{\sin(\omega \tau_2)\cos(\omega \tau_1)}{k_1} = 0
\end{equation}
for antisymmetric modes, or
\begin{equation}
\label{eq:toy_model_discriminant2}
-\frac{\sin(\omega \tau_1)\sin(\omega \tau_2)}{k_2} + 
\frac{\cos(\omega \tau_2)\cos(\omega \tau_1)}{k_1} = 0
\end{equation}
for symmetric modes, where $\tau_i = x_i/c_i$.

In the simple case where $c_1=c_2$, the solutions are
\begin{equation}
\omega_k = \frac{k\pi}{\tauT} \qquad \mbox{or} \qquad
\omega_k = \frac{\left(k+\frac{1}{2}\right)\pi}{\tauT}
\label{eq:toy_model_zero}
\end{equation}
for odd and even modes respectively, and where $\tauT = \tau_1 + \tau_2$.

\begin{figure}[htbp]
\includegraphics[width=\columnwidth]{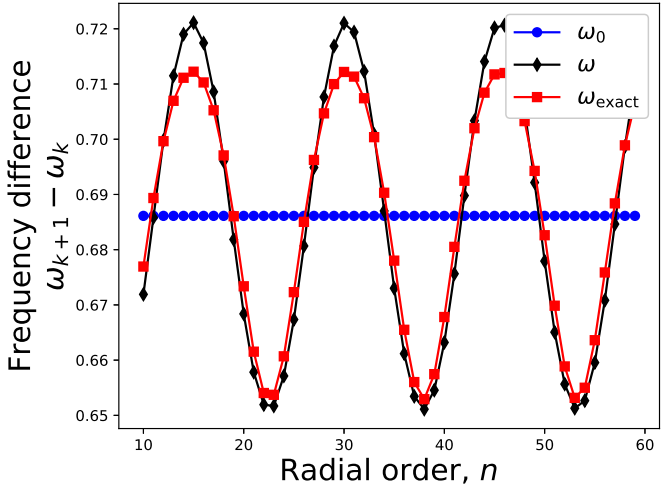}
\caption{Large frequency separations for various calculations of the frequency:
$\omega_0$ corresponds to the zeroth order expression (i.e. no perturbations to
the sound velocity are included), $\omega$ to the first order approximation, and
$\omega_{\mathrm{exact}}$ to exact solutions of the discriminant equations.
\label{fig:toy_model_delta}}
\end{figure}

When $c_1$ and $c_2$ differ, we can perform a first order perturbative analysis
by introducing a small parameter $\epsilon$ as follows:
\begin{equation}
k_2 = k_1 (1 + \epsilon).
\end{equation}
This leads to the following corrections on odd and even modes respectively:
\begin{eqnarray}
\delta \omega &=& \frac{(-1)^k}{2\tauT} \epsilon \sin(\omega_0 (\tau_1-\tau_2)), \\
\delta \omega &=& \frac{(-1)^k}{2\tauT} \epsilon \cos(\omega_0 (\tau_1-\tau_2)),
\end{eqnarray}
where $\omega_0$ corresponds to the unperturbed frequencies given in
Eq.~(\ref{eq:toy_model_zero}).  Combining the even and odd cases and
including both the zeroth and first order components yields:
\begin{equation}
\label{eq:toy_model_frequencies}
\omega_n = \frac{1}{2\tauT} \left[ n\pi + \epsilon \sin\left(n\pi\frac{\tau_1}{\tauT}\right) \right],
\end{equation}
where odd values of $n$ correspond to even solutions and vice versa. The period
of the frequency perturbation is analogous but somewhat simplified compared to
the more general formula given in \citet{Monteiro1994} for non-rotating stars.
Figure~\ref{fig:toy_model_delta} compares large frequency separations using the
first order expression above (Eq.~(\ref{eq:toy_model_frequencies})) and those
obtained from exact solutions to the discriminants given in
Eqs.~(\ref{eq:toy_model_discriminant1}) and~(\ref{eq:toy_model_discriminant2})
for of the values $\tau_1$ and $\tauT$ from model \texttt{M7}, the most extreme
case.  As can be seen, the first order expression gives an accurate idea of the
period of the frequency deviation, and a rough idea of its amplitude.

\section{Wave refraction and reflection}
\label{app:refraction}

In this section, we recall some of the basic principles behind the
Snell-Descartes law including partial wave reflection.  A more complete
treatment can be found in various textbooks such as \citet{Brekhovskikh1980}.

We begin with a simple plane-parallel model using Cartesian coordinates. 
This can also be thought of as a local approximation to a more complex system. 
A discontinuity in density is located at $z=0$.  The media below and above this
discontinuity is assumed to be uniform.  Under these conditions, the fluid
dynamic equations take on the following expressions:
\begin{eqnarray}
0 &=& \drho + \div \vect{\xi}, \\
\rho_0 \ddpart{\vect{\xi}}{t} &=& -P_0 \grad \dP, \\
\dP &=& \Gamma_1 \drho.
\end{eqnarray}
The interface conditions, as explained in App.~\ref{app:interface_conditions},
ensure the continuity of $\dP$ and $\xi_z$.  Combining these equations leads to:
\begin{equation}
\ddpart{}{t}\left(\dP\right) = c_0^2 \lapl \left( \dP \right).
\end{equation}

Because of the partial reflection at the boundary, we cannot consider a
plane-parallel wave in isolation but have to include the reflected wave.  For
the sake of generality, we consider such a combination both above and below the
discontinuity.  The leads to following generic solution:
\begin{equation}
\left(\dP\right)^{\pm} = A_1^{\pm} \exp(i \vect{k}_1^{\pm} \cdot \vect{x} + i\omega t)
                       + A_2^{\pm} \exp(i \vect{k}_2^{\pm} \cdot \vect{x} + i\omega t),
\end{equation}
where the superscripts `$+$' and `$-$' designate the upper and lower domains,
respectively.  The standing wave equivalent to the above solution would take on
the expression:
\begin{equation}
\left(\dP\right)^{\pm} = \left[A_1^{\pm} \cos(\vect{k}_1^{\pm} \cdot \vect{x})
                       + A_2^{\pm} \cos(\vect{k}_2^{\pm} \cdot \vect{x}) \right]
                       \exp(i\omega t).
\end{equation}
The wave vectors take on the following form:
\begin{equation}
\vect{k}_1^{\pm} = \vkpar + k_z^{\pm} \ez, \qquad
\vect{k}_2^{\pm} = \vkpar - k_z^{\pm} \ez.
\end{equation}
The horizontal wave vector, $\vkpar$, is preserved between the two domains as a
result of the continuity of the horizontal gradient of $\dP$ at the
discontinuity.  When combined with the dispersion relation, this leads to
Snell-Descartes' law.

The continuity of $\dPi$ leads to the relation:
\begin{equation}
A_1^+ + A_2^+ = A_1^- + A_2^-.
\end{equation}
The continuity of $\xi_z$ leads to:
\begin{equation}
\frac{A_1^+ - A_2^+}{\rho_0^+} = \frac{A_1^- - A_2^-}{\rho_0^-}.
\end{equation}
Combining these two equations leads the following matrix relation between
the amplitudes:
\begin{equation}
\left[
\begin{array}{c}
A_1^+ \\
A_2^+
\end{array}
\right]
=
\frac{1}{2}
\left[
\begin{array}{cc}
1 + \eta & 1 - \eta \\
1 - \eta & 1 + \eta
\end{array}
\right]
\left[
\begin{array}{c}
A_1^- \\
A_2^-
\end{array}
\right],
\end{equation}
where
\begin{equation}
\eta = \frac{\rho_0^+}{\rho_0^-} \frac{k_z^-}{k_z^+}.
\end{equation}
When the wave vector is nearly perpendicular to the discontinuity, that is, $\kpar
\ll k_z$, $\eta$ takes on the following approximate expression:
\begin{equation}
\eta \simeq \sqrt{\frac{\rho_0^+}{\rho_0^-}} = \frac{c_0^-}{c_0^+}.
\end{equation}

\end{document}